\documentclass[a4paper, times, 10pt, twocolumn, twoside,top=3.5cm,bottom=3.5cm,left=2.5cm,right=2cm]{article}

\usepackage{ISARC}

\usepackage{lscape}
\usepackage{hologo}
\usepackage{gensymb}

\newtheorem{assumption}{Assumption}

\begin{document}

\linespread{0.5}

\title{Modeling and control of 5-DoF boom crane}

\author{Michele Ambrosino$^{a}$ Marc Berneman$^{b}$ Gianluca Carbone$^{a}$ \\
\scalebox{1.2}{\textbf{Rémi Crépin$^{a}$ Arnaud Dawans$^c$ Emanuele Garone$^{a}$}}
}

\affiliation{
$^a$Service d’Automatique et d’Analyse des Systèmes, Université Libre de Bruxelles, Brussels, Belgium\\
\medskip
$^b$ Vrije Universiteit Brussel,$^c$Entreprises Jacques Delens S.A., Brussels, Belgium \\
\smallskip
\href{mailto:e.author1@aa.bb.edu}{Michele.Ambrosino@ulb.ac.be}, 
\href{mailto:e.author1@aa.bb.edu}{marc.berneman@vub.be},
\href{mailto:e.author1@aa.bb.edu}{Gianluca.Carbone@ulb.ac.be}, \\
\href{mailto:e.author1@aa.bb.edu}{Remi.Crepin@ulb.ac.be},
\href{mailto:e.author1@aa.bb.edu}{adawans@jacquesdelens.be},
\href{mailto:e.author1@aa.bb.edu}{egarone@ulb.ac.be}
}

\maketitle 
\thispagestyle{fancy} 
\pagestyle{fancy}

\begin{abstract}
Automation of cranes can have a direct impact on the productivity of construction projects. In this paper, we focus on the control of one of the most used cranes, the boom crane. Tower cranes and overhead cranes have been widely studied in the literature, whereas the control of boom cranes has been investigated only by a few works. Typically, these works make use of simple models making use of a large number of simplifying assumptions (e.g. fixed length cable, assuming certain dynamics are uncoupled, etc.)
A first result of this paper is to present a fairly complete nonlinear dynamic model of a boom crane taking into account all coupling dynamics and where the only simplifying assumption is that the cable is considered as rigid. The boom crane involves pitching and rotational movements, which generate complicated centrifugal forces, and consequently, equations of motion highly nonlinear.
On the basis of this model, a control law has been developed able to perform position control of the crane while actively damping the oscillations of the load.
The effectiveness of the approach has been tested in simulation with 
realistic physical parameters and tested in the presence of wind disturbances. 
\end{abstract}


\section{Introduction}\label{sec:Introduction}

A crane is a type of machine, generally equipped with a hoist rope, that is used to move materials. Cranes can be classified in overhead cranes \cite{25}-\cite{26}, offshore cranes \cite{27}–\cite{30}, and rotary cranes \cite{31}-\cite{kncntr}. Currently, the automation of cranes is still in a relatively early phase. To improve the efficiency and safety of cranes some control approaches have be proposed using sliding-mode control \cite{33}-\cite{33b}, optimal control \cite{34}, adaptive control \cite{39}, prediction control \cite{44}, intelligent control \cite{45}.
\medskip

In this paper we focus on the modeling and control of a very common type of rotary crane, know as 'boom crane'.

\medskip

Compared with other cranes, boom cranes have higher flexibility and lower energy consumption. Therefore, boom cranes have been widely used in the maintenance of buildings and to handle masonry in urban streets and construction sites. There the cranes have a boom that can rotate in two directions (e.g. pitch and yaw motions) and the load swing can be split into two dimensions. Consequently, the nonlinear dynamic models of boom cranes are more complex than those of other types of cranes.

\medskip

In recent years, a number of studies have been carried out to solve the control problems of such complex systems. \cite{46}-\cite{47} proposed the use of S-curve trajectories as an open-loop control approaches to achieve anti-sway control for the payload. Moreover, input shaping has been widely applied to control boom cranes \cite{49}. However, the open-loop control strategies are sensitive to external disturbances and to model mismatch. Motivated by these reasons, closed-loop control approaches have been proposed. In \cite{51} the combination of command shaping and feedback control was proposed which can reduce payload oscillation. In \cite{56}, a state feedback control law based on linearized model is used to achieve the control objectives. In \cite{yang2017}-\cite{sun2017} the authors proposed a Proportional-Derivative (PD) controller with gravity compensation based on the nonlinear model of the boom crane. In \cite{erg} the authors present constrained control for boom cranes.

\medskip

Most of the existing closed-loop control laws for boom cranes have two main drawbacks: 
\begin{enumerate}
    \item The dynamic of the hoisting mechanism is neglected (e.g. the length of the cable is considered as constant).
    \item In the design of the control law, the possibility of measuring the oscillations of the payload (e.g. angular positions and speeds) is usually ignored.
\end{enumerate}

In order to address these problems, we propose a control law that exploits all states of the system to control it. The proposed control scheme is based on a detailed mathematical model in which we takes into account all the degrees of freedom (DoFs) that characterize this type of system (i.e. the two rotations, the length of the rope, and the payload swing angles). Realistic physical parameters of an existing boom crane are used in simulation tests to show the effectiveness of the proposed approach.

\section{Dynamic Model}

\begin{figure}[ht!]
\centering
\includegraphics[width=0.9\columnwidth]{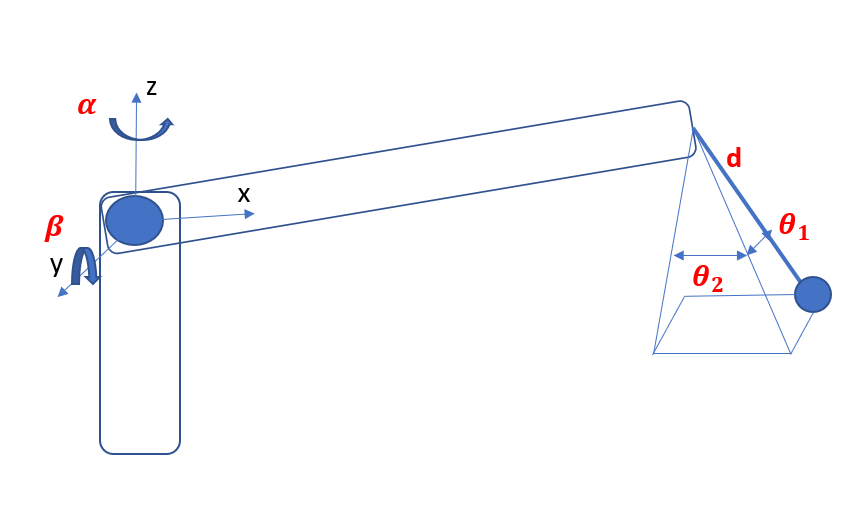}
\caption{\label{fig:crane} Model of a boom crane}
\end{figure}

The type of crane considered in this paper (see Fig.\ref{fig:crane}) is represented by five generalized coordinates: $\alpha$ is the slew angle of the tower, $\beta$ is the luff angle of the boom, d is the length of the rope, $\theta_1$ is the tangential pendulation due to the motion of the tower, and $\theta_2$ is the radial sway due to the motion of the boom. 
\newline 
The equations of the motion obtained using the Euler-Lagrange approach are
\begin{equation}\label{eq:al}
    \begin{split}
        I_t\ddot{\alpha} + d^2\ddot{\alpha}m + \ddot{\alpha}l_B^2mC_{\beta}^2 + (\ddot{\alpha}l_B^2m_BC_{\beta}^2)/4 - d^2\ddot{\theta_2}mS_{\theta_3} \\ + 2\dot{d}d\dot{\alpha}m - 2d^2\dot{\theta_1}\dot{\theta_2}mC_{\theta_1} - d^2\ddot{\alpha}mC_{\theta_1}^2C_{\theta_4}^2 - \dot{\alpha}\dot{\beta}l_B^2mS_{2\beta} \\- (\dot{\alpha}\dot{\beta}l_B^2m_BS_{2\beta})/4 - 2\dot{d}d\dot{\theta_2}mS_{\theta_3} + \ddot{d}l_BmC_{\beta}C_{\theta_4}S_{\theta_3} \\- 2\dot{d}d\dot{\alpha}mC_{\theta_1}^2C_{\theta_4}^2 + 2d^2\dot{\theta_1}\dot{\theta_2}mC_{\theta_1}C_{\theta_4}^2 + d^2\ddot{\theta_1}mC_{\theta_1}C_{\theta_4}S_{\theta_4} \\ + 2d\ddot{\alpha}l_BmC_{\beta}S_{\theta_4} + 2\dot{d}\dot{\alpha}l_BmC_{\beta}S_{\theta_4} - d^2\dot{\theta_1}^2mC_{\theta_4}S_{\theta_3}S_{\theta_4}\\ + 2d\dot{\alpha}\dot{\theta_2}l_BmC_{\beta}C_{\theta_4} - 2d\dot{\alpha}\dot{\beta}l_BmS_{\beta}S_{\theta_4} \\+ 2d^2\dot{\alpha}\dot{\theta_2}mC_{\theta_1}^2C_{\theta_4}S_{\theta_4} + d\ddot{\theta_1}l_BmC_{\beta}C_{\theta_1}C_{\theta_4} \\+ 2\dot{d}\dot{\theta_1}l_BmC_{\beta}C_{\theta_1}C_{\theta_4} + 2\dot{d}d\dot{\theta_1}mC_{\theta_1}C_{\theta_4}S_{\theta_4} \\+ d\ddot{\beta}l_BmC_{\theta_4}S_{\beta}S_{\theta_3} - d\ddot{\theta_2}l_BmC_{\beta}S_{\theta_3}S_{\theta_4} \\- 2\dot{d}\dot{\theta_2}l_BmC_{\beta}S_{\theta_3}S_{\theta_4} + d\dot{\beta}^2l_BmC_{\beta}C_{\theta_4}S_{\theta_3}\\ - d\dot{\theta_1}^2l_BmC_{\beta}C_{\theta_4}S_{\theta_3} +2d^2\dot{\alpha}\dot{\theta_1}mC_{\theta_1}C_{\theta_4}^2S_{\theta_3} \\- d\dot{\theta_2}^2l_BmC_{\beta}C_{\theta_4}S_{\theta_3} - 2d\dot{\theta_1}\dot{\theta_2}l_BmC_{\beta}C_{\theta_1}S_{\theta_4} = u_1,
    \end{split}
\end{equation}

\begin{equation}\label{eq:bt}
    \begin{split}
        I_B\ddot{\beta} + \ddot{\beta}l_B^2m + (\ddot{\beta}l_B^2m_B)/4 + gl_BmC_{\beta} + (gl_Bm_BC_{\beta})/2 \\+ (\dot{\alpha}^2l_B^2mS_{2\beta})/2 + (\dot{\alpha}^2l_B^2m_BS_{2\beta})/8 - \ddot{d}l_BmS_{\beta}S_{\theta_2} \\- \ddot{d}l_BmC_{\beta}C_{\theta_1}C_{\theta_2} + d\dot{\alpha}^2l_BmS_{\beta}S_{\theta_2} + d\dot{\theta_2}^2l_BmS_{\beta}S_{\theta_2} \\ - d\ddot{\theta_2}l_BmC_{\theta_2}S_{\beta} - 2\dot{d}\dot{\theta_2}l_BmC_{\theta_2}S_{\beta} + d\ddot{\theta_1}l_BmC_{\beta}C_{\theta_2}S_{\theta_1} \\+ d\ddot{\theta_2}l_BmC_{\beta}C_{\theta_1}S_{\theta_2} + 2\dot{d}\dot{\theta_1}l_BmC_{\beta}C_{\theta_2}S_{\theta_1} \\+ 2\dot{d}\dot{\theta_2}l_BmC_{\beta}C_{\theta_1}S_{\theta_2} + d\ddot{\alpha}l_BmC_{\theta_2}S_{\beta}S_{\theta_1} \\+ 2\dot{d}\dot{\alpha}l_BmC_{\theta_2}S_{\beta}S_{\theta_1} + d\dot{\theta_1}^2l_BmC_{\beta}C_{\theta_1}C_{\theta_2} \\+ d\dot{\theta_2}^2l_BmC_{\beta}C_{\theta_1}C_{\theta_2} - 2d\dot{\theta_1}\dot{\theta_2}l_BmC_{\beta}S_{\theta_1}S_{\theta_2} \\- 2d\dot{\alpha}\dot{\theta_2}l_BmS_{\beta}S_{\theta_1}S_{\theta_2} + 2d\dot{\alpha}\dot{\theta_1}l_BmC_{\theta_1}C_{\theta_2}S_{\beta} = u_2,
    \end{split}
\end{equation}

\begin{equation}\label{eq:d}
    \begin{split}
        \ddot{d}m - d\dot{\alpha}^2m - d\dot{\theta_2}^2m - d\dot{\theta_1}^2mC_{\theta_2}^2 - gmC_{\theta_1}C_{\theta_2} \\ + d\dot{\alpha}^2mC_{\theta_1}^2C_{\theta_2}^2 - \ddot{\beta}l_BmS_{\beta}S_{\theta_2} - \dot{\alpha}^2l_BmC_{\beta}S_{\theta_2} \\ - \dot{\beta}^2l_BmC_{\beta}S_{\theta_2} + 2d\dot{\alpha}\dot{\theta_2}mS_{\theta_1} - \ddot{\beta}l_BmC_{\beta}C_{\theta_1}C_{\theta_2} \\+ \ddot{\alpha}l_BmC_{\beta}C_{\theta_2}S_{\theta_1} + \dot{\beta}^2l_BmC_{\theta_1}C_{\theta_2}S_{\beta} \\- 2d\dot{\alpha}\dot{\theta_1}mC_{\theta_1}C_{\theta_2}S_{\theta_2} - 2\dot{\alpha}\dot{\beta}l_BmC_{\theta_2}S_{\beta}S_{\theta_1} = u_3,
    \end{split}
\end{equation}

\begin{equation} \label{eq:th1}
    \begin{split}
        dmC_{\theta_2}(gS_{\theta_1} + 2\dot{d}\dot{\theta_1}C_{\theta_2} + d\ddot{\theta_1}C_{\theta_2} - \dot{\beta}^2l_BS_{\beta}S_{\theta_1} \\- 2d\dot{\theta_1}\dot{\theta_2}S_{\theta_2} + \ddot{\alpha}l_BC_{\beta}C_{\theta_1} + d\ddot{\alpha}C_{\theta_1}S_{\theta_2} + 2\dot{d}\dot{\alpha}C_{\theta_1}S_{\theta_2}\\ + \ddot{\beta}l_BC_{\beta}S_{\theta_1} - d\dot{\alpha}^2C_{\theta_1}C_{\theta_2}S_{\theta_1} + 2d\dot{\alpha}\dot{\theta_2}C_{\theta_1}C_{\theta_2} \\- 2\dot{\alpha}\dot{\beta}l_BC_{\theta_1}S_{\beta}) = 0,    \end{split}
\end{equation}

\begin{equation} \label{eq:th2}
    \begin{split}
        -dm(d\ddot{\alpha}S_{\theta_1} - 2\dot{d}\dot{\theta_2} - d\ddot{\theta_2} + 2\dot{d}\dot{\alpha}S_{\theta_1} - gC_{\theta_1}S_{\theta_2} \\ - (d\dot{\theta_1}^2S_{2\theta_2})/2 + \dot{\alpha}^2l_BC_{\beta}C_{\theta_2} + \dot{\beta}^2l_BC_{\beta}C_{\theta_2}\\ + \ddot{\beta}l_BC_{\theta_2}S_{\beta} + \dot{\beta}^2l_BC_{\theta_1}S_{\beta}S_{\theta_2} + d\dot{\alpha}^2C_{\theta_1}^2C_{\theta_2}S_{\theta_2}\\ + 2d\dot{\alpha}\dot{\theta_1}C_{\theta_1}C_{\theta_2}^2 - \ddot{\beta}l_BC_{\beta}C_{\theta_1}S_{\theta_2} + \ddot{\alpha}l_BC_{\beta}S_{\theta_1}S_{\theta_2}\\ - 2\dot{\alpha}\dot{\beta}l_BS_{\beta}S_{\theta_1}S_{\theta_2}) = 0. 
    \end{split}
\end{equation}

where m, $m_b$ denote the load mass, and the boom mass, respectively, $l_b$ is the boom length, $I_t$ is the inertia moment of the tower, and $I_b$ is the inertia moment of the boom. Moreover, the following abbreviations are used:
\newline
$S_{\alpha}\triangleq sin(\alpha)$, $S_{\beta}\triangleq sin(\beta), S_{\theta_1}\triangleq sin(\theta_1)$,$S_{\theta_2} \triangleq sin(\theta_2), C_{\alpha} \triangleq cos(\alpha), C_{\beta} \triangleq cos(\beta), C_{\theta_1} \triangleq cos(\theta_1), C_{\theta_2} \triangleq cos(\theta_2)$.

\medskip

The system dynamics (\ref{eq:al})-(\ref{eq:th2}) can be rewritten in matrix form as  

\begin{equation}\label{eq:modelmatrix}
{{M(q)\ddot{q} + C(q,\dot{q})\dot{q} + G(q)} = {\begin{bmatrix}
I_{3x3} \\ 0_{2x2} \end{bmatrix}}u, }
\end{equation}

where q = $[\alpha, \beta, d, \theta_1, \theta_2]^T \in{\mathbb{R}^5}$ represents the state vector, and u = $[u_1, u_2, u_3]^T \in{\mathbb{R}^3}$ is the control input vector.
The matrices M(q) $\in{\mathbb{R}^{5x5}}$, $ C(q,\dot{q})\in{\mathbb{R}^{5x5}}$, and G(q) $\in{\mathbb{R}^{5}}$ represent the inertia matrix, centripetal-Coriolis forces, and gravity term, respectively. 





As one can seen from (\ref{eq:modelmatrix}), the boom crane is an underactuated system, having fewer independent actuators than system degrees of freedom (DoFs). Thus, we can rewrite its model as

\begin{equation}\label{eq:eq_ac}
\begin{split}
   M_{11}(q)\ddot{q}_1+M_{12}(q)\ddot{q}_2\\+C_{11}(q,\dot{q})\dot{q}_1+C_{12}(q,\dot{q})\dot{q}_2 + G_1(q) = U,
\end{split}
\end{equation}

\begin{equation}\label{eq:eq_na}
\begin{split}
   M_{21}(q)\ddot{q}_1+M_{22}(q)\ddot{q}_2+C_{21}(q,\dot{q})\dot{q}_1\\+C_{22}(q,\dot{q})\dot{q}_2 + G_2(q) = 0,
\end{split}
\end{equation}

where $q_1 = [\alpha \quad \beta\quad d]^T $ is the vector of actuated states and $q_2 = [\theta_1 \quad \theta_2]^T $ of non-actuated states and

\begin{equation*}\label{eq:m11}
M_{11}(q) = \begin{bmatrix}
m_{11}&m_{12}&m_{13}\\
m_{21}&m_{22}&m_{23}\\
m_{31}&m_{32}&m_{33}\\
\end{bmatrix},
M_{12}(q) = \begin{bmatrix}
m_{14}&m_{15}\\
m_{24}&m_{25}\\
\end{bmatrix},
\end{equation*}

\begin{equation*}
M_{21}(q) = \begin{bmatrix}
m_{41}&m_{42}&m_{43}\\
m_{51}&m_{52}&m_{53}\\
\end{bmatrix},
M_{22}(q) = \begin{bmatrix}
m_{44}&0\\
0&m_{55}\\
\end{bmatrix},
\end{equation*}

\begin{equation*}\label{eq:c11}
C_{11}(q,\dot{q}) = \begin{bmatrix}
c_{11}&c_{12}&c_{13}\\
c_{21}&c_{22}&c_{23}\\
c_{31}&c_{32}&c_{33}\\
\end{bmatrix},
C_{12}(q,\dot{q}) = \begin{bmatrix}
c_{14}&c_{15}\\
c_{24}&c_{25}\\
\end{bmatrix},
\end{equation*}

\begin{equation*}
C_{21}(q,\dot{q}) = \begin{bmatrix}
c_{41}&c_{42}&c_{43}\\
c_{51}&c_{52}&c_{53}\\
\end{bmatrix},
C_{22}(q,\dot{q}) = \begin{bmatrix}
c_{44}&0\\
0&c_{55}\\
\end{bmatrix},
\end{equation*}

\begin{equation*}
    G_1(q) = \begin{bmatrix}
    0 \\
    g_2\\
    g_3\\
    \end{bmatrix},
    G_2(q) = \begin{bmatrix}
    g_4\\
    g_5\\
    \end{bmatrix},
    U = \begin{bmatrix}
    u_1 \\
    u_2\\
    u_3\\
    \end{bmatrix}.
\end{equation*}

\section{Control Design}\label{sec:cntrl}

The aim of the control is to move the crane to the desired position and to dampen the swing angles of the load. In our development we will consider the following reasonable assumptions.

\begin{assumption}\label{ass:1}
The payload swing are such that $\lvert \theta_{1,2} \rvert\ <{\frac{\pi}{2}}.$
\end{assumption}

\begin{assumption}\label{ass:2}
The cable length is always greater than zero to avoid singularity in the model (\ref{eq:modelmatrix}), i.e. $d(t) > 0, \forall t\geq0.$ \end{assumption}

\medskip

As one can see, (\ref{eq:eq_na}) can be rewritten as

\begin{equation}\label{eq:part_na}
\begin{split}
    \ddot{q}_2 = - M_{22}^{-1}(q)(M_{21}(q)\ddot{q}_1+C_{21}(q,\dot{q})\dot{q}_1\\+C_{22}(q,\dot{q})\dot{q}_2+G_2(q)).
\end{split}
\end{equation}

It is worth noticing that in (\ref{eq:part_na}) the $M_{22}(q)$ is a positive definite matrix due to Assumptions (\ref{ass:1})-(\ref{ass:2}). 

\medskip

Substituting (\ref{eq:part_na}) into (\ref{eq:eq_ac}), one obtains

\begin{equation}\label{eq:bar}
    {\bar M(q)\ddot{q}_1 + \bar C_{1}(q,\dot{q})\dot{q}_1 + \bar C_{2}(q,\dot{q})\dot{q}_2 + \bar G(q) = U }, 
\end{equation}

where

\begin{align*}
    \bar M(q) = M_{11}(q)-M_{12}(q)M_{22}^{-1}(q)M_{21}(q), \quad\quad\quad\:\:\: \\
    \bar C_{1}(q,\dot{q}) = C_{11}(q,\dot{q}) - M_{12}(q)M_{22}^{-1}(q)C_{12}(q,\dot{q}),\quad\\
    \bar C_{2}(q,\dot{q}) = C_{12}(q,\dot{q}) - M_{12}(q)M_{22}^{-1}(q)C_{22}(q,\dot{q}),\quad\\
    \bar G(q) = G_1(q) - M_{12}(q)M_{22}^{-1}(q)G_2(q).\quad\quad\quad\quad
\end{align*}

According to Assumptions (\ref{ass:1})-(\ref{ass:2}), the matrix $\bar M$ is positive definite. Then, (\ref{eq:bar}) can be rewritten as

\begin{equation}\label{eq:par_a}
    {\ddot{q}_1 = \bar M^{-1}(q)(U - \bar C_{1}(q,\dot{q})\dot{q}_1 - \bar C_{2}(q,\dot{q})\dot{q}_2 - \bar G(q))}
\end{equation}

Substituting (\ref{eq:par_a}) into (\ref{eq:part_na}) yields

\begin{equation}\label{eq:part_na2}
\begin{split}
    \ddot{q}_2 = - M_{22}^{-1}(q)(M_{21}(q)\bar M^{-1}(q)( - \bar C_{1}(q,\dot{q})\dot{q}_1  - \bar C_{2}(q,\dot{q})\dot{q}_2 \\- \bar G(q)+U)+C_{21}(q,\dot{q})\dot{q}_1+C_{22}(q,\dot{q})\dot{q}_2+G_2(q)).
\end{split}
\end{equation}

Following the classical approach of a feedback linearization technique, (\ref{eq:par_a}) can be “linearized”  by using the control law

\begin{equation}\label{eq:U}
    {U = \bar M(q)v + \bar C_{1}(q,\dot{q})\dot{q}_1 + \bar C_{2}(q,\dot{q})\dot{q}_2 + \bar G(q)}. 
\end{equation}

Thus, (\ref{eq:par_a}) becomes

\begin{equation}\label{eq:dq_p}
    {\ddot{q}_1 = v,}
\end{equation}

where $v \in\mathbb{R}^{3}$ as additional control inputs.

\medskip

To move the crane to the desired position, the additional control inputs (\ref{eq:dq_p}) can be chosen as

\begin{equation}\label{eq:va}
    {v = \ddot{q}_{1d}-K_{ad}(\dot{q}_1-\dot{q}_{1d})-K_{ap}({q_1}+{q_{1d}})},
\end{equation}

where $K_{ad} = diag(K_{ad1},K_{ad2},K_{ad3})$, $K_{ap} = diag(K_{ap1},K_{ap2},K_{ap3})$ are positive diagonal matrices.
\newline
Substituting (\ref{eq:va}) into (\ref{eq:dq_p}), we obtain
\begin{equation}\label{eq:dq_t}
    {\ddot{\tilde{q}}_1 + K_{ad}\dot{\tilde{q}}_1 + K_{ap}\tilde{q}_1 = 0},
\end{equation}

where $\tilde{q} = q_1 - q_{1d}$ is the tracking error vector of the actuated states. (\ref{eq:dq_t}) is exponentially stable for every $K_{ad} > 0$ and $K_{ap} > 0$.

\medskip

To stabilize the non-actuated states $q_2$, following what is proposed in \cite{partial}, we define a second additional inputs as

\begin{equation}\label{eq:vu}
    {v_{u}=-K_{ud}\dot{q}_2-K_{up}q_2},
\end{equation}

where $v_u \in\mathbb{R}^2$ are additional inputs which take into account the non-actuated states, $K_{up} = diag(K_{up1},K_{up2})$, $K_{up} = diag(K_{ud1},K_{ud2})$ are positive diagonal matrices.

\medskip

Considering $q_{1d} = const$, the overall additional inputs are proposed by linearly combining (\ref{eq:va}) and (\ref{eq:vu})

\begin{equation}\label{eq:V}
\begin{split}
    v = -K_{ad}\dot{q}_1-K_{ap}(q_1-q_{1d})-\alpha(K_{ud}\dot{q}_2+K_{up}q_2),
\end{split}
\end{equation}

where 

\begin{equation}
 \alpha = \begin{bmatrix}
 \alpha_1&0\\
 0&\alpha_2\\
 0&0\\
 \end{bmatrix}
\end{equation}

is a weighting matrix.

Substituting the (\ref{eq:V}) into (\ref{eq:U}) the overall control law is obtained as

\begin{equation}\label{eq:U2}
\begin{split}
    U = (\bar C_{1}(q,\dot{q}) - \bar M(q)K_{ad})\dot{q}_1 + (\bar C_{2}(q,\dot{q})-\bar M(q)\alpha K_{ud})\dot{q}_2 \\ -\bar M(q)K_{ap}(q_1-q_{1d})-\bar M(q)\alpha K_{up}q_2  + \bar G(q).
\end{split}
\end{equation}

Replacing (\ref{eq:U2}) into (\ref{eq:part_na2}), we obtain 

\begin{equation}\label{eq:partial_na3}
    \begin{split}
        \ddot{q}_2 = - M_{22}^{-1}(q)(-M_{21}(q)(K_{ad}\dot{q}_1+K_{ap}q_1 \\ +\alpha(K_{ud}\dot{q}_2+K_{up}q_2)) +C_{21}(q,\dot{q})\dot{q}_1+C_{22}(q,\dot{q})\dot{q}_2+G_2(q)).
    \end{split}
\end{equation}

Considering Assumption 1, in the rest of this Section we have to demonstrate that  (\ref{eq:partial_na3}) converges to the equilibrium point expressed by: $q_2 = \dot{q}_2 = 0$ to achieve the control goal.

\medskip

Setting $q_1 = q_{1d}$ in (\ref{eq:partial_na3}), one achieves

\begin{equation}\label{eq:partial_na4}
    \begin{split}
        \ddot{q}_2 = - M_{22}^{-1}(q)(-M_{21}(q)(\alpha(K_{ud}\dot{q}_2+K_{up}q_2)) \\+C_{22}(q,\dot{q})\dot{q}_2+G_2(q)).
    \end{split}
\end{equation}

The stability analysis of (\ref{eq:partial_na4}) 
is analyzed by linearizing (\ref{eq:partial_na4}) around the equilibrium point $q_2=\dot{q}_2=0$.
\newline
We can rewrite (\ref{eq:partial_na4}) as

\begin{align*}
    z_1 = \theta_1, \quad z_2 = \dot{\theta}_1, \quad z_3 = \theta_2, \quad z_4 = \dot{\theta}_2. 
\end{align*}

Then, we obtain the following state-space forms:

\begin{equation}\label{eq:z1}
    {\dot{z}_1 = z_2},
\end{equation}
\begin{equation}\label{eq:z2}
    {\dot{z}_2 = h_1(z)},
\end{equation}\label{eq:z3}
\begin{equation}
    {\dot{z}_3 = z_4},
\end{equation}
\begin{equation}\label{eq:z4}
    {\dot{z}_4 = h_2(z)},
\end{equation}

with $z = [z_1\quad z_2 \quad z_3 \quad z_4]^T$ as a state vector. Linearizing (\ref{eq:z1})-(\ref{eq:z4}) around $z=0$, we obtain 

\begin{equation}\label{eq:lin_sys}
    {\dot{z}=Az,}
\end{equation}

where 
 
 \begin{equation}\label{eq:A}
 \begin{split}
     A = \begin{bmatrix}
     0&1&0&0\\
     \frac{\partial h_1}{\partial z_1}&\frac{\partial h_1}{\partial z_2}&\frac{\partial h_1}{\partial z_3}&\frac{\partial h_1}{\partial z_4}\\
     0&0&0&1\\
     \frac{\partial h_2}{\partial z_1}&\frac{\partial h_2}{\partial z_2}&\frac{\partial h_2}{\partial z_3}&\frac{\partial h_2}{\partial z_4}\\
     \end{bmatrix}_{z=0}
     \\
     = \begin{bmatrix}
     0&1&0&0\\
     a_{11}&a_{12}&0&0\\
     0&0&0&1\\
     0&0&a_{21}&a_{22}\\
     \end{bmatrix}.
      \end{split}
 \end{equation}
 
 The non-zero elements in (\ref{eq:A}) are the flowwing:
 
 \begin{equation}\label{eq:a11}
     a_{11} = -\frac{(g - a_1K_{pu1}l_Bcos\beta)}{d},
 \end{equation}
\begin{equation}\label{eq:a12}
     a_{12} = \frac{a_1K_{du1}l_Bcos\beta}{d},
 \end{equation}
   \begin{equation}\label{eq:a21}
     a_{21} = -\frac{(g + a_2K_{pu2}l_Bsin\beta)}{d},
 \end{equation}
  \begin{equation}\label{eq:a22}
     a_{22} = -\frac{a_2K_{du2}l_Bsin\beta}{d}.
 \end{equation}
 
The linearized system (\ref{eq:lin_sys}) is stable around the equilibrium point $z=0$, if the A matrix (\ref{eq:A}) is a Hurwiz matrix. Therefore, it is necessary to properly choose the control parameters that appear in (\ref{eq:a11})-(\ref{eq:a22}). In this way, (\ref{eq:lin_sys}) is stable around equilibrium point z = 0, which leads to the local stability
of (\ref{eq:partial_na3}). In the Section \ref{sec:Sim} the values for each of the control parameters are listed. 
 
\section{Simulation Results}\label{sec:Sim}

In this section, three different simulation scenarios will be shown to demonstrate the proposed control scheme. In each of them the goal is to move the crane to a desired position and to reduce the swings of the payload as much as possible. In the second and third simulation, the effects of a gust of wind for the payload will be shown. 

\medskip

To get realistic values for the simulation tests, we consider a small boom crane: the NK 1000 Mini Crane from NEMAASKO \cite{NK100_user_manual}. Some parameters are taken directly from the datasheets. Others, like the boom dimensions were estimated by CAD simulations.
\newline
The crane system parameters are selected as follows:

\begin{align*}
    I_t = 207.13kgm^2,\quad l_B=6.2m, \quad m_B=312.2kg, \\ I_B=2068kgm^2, \quad  g=9.81ms^{-2}, \quad  m=50kg. \;\;\;
\end{align*}

The control parameters for
controller (\ref{eq:V}) are set as $K_{ad}=diag(100,100,150)$, $K_{ap}=diag(10,20,50)$, $K_{ud}=diag(120,120)$, $K_{up}=diag(10,10)$, $a_1=-1$, $a_2=sign(\beta)$. 

\begin{figure}[ht!]
\centering
\includegraphics[width=8cm, height=2.5cm]{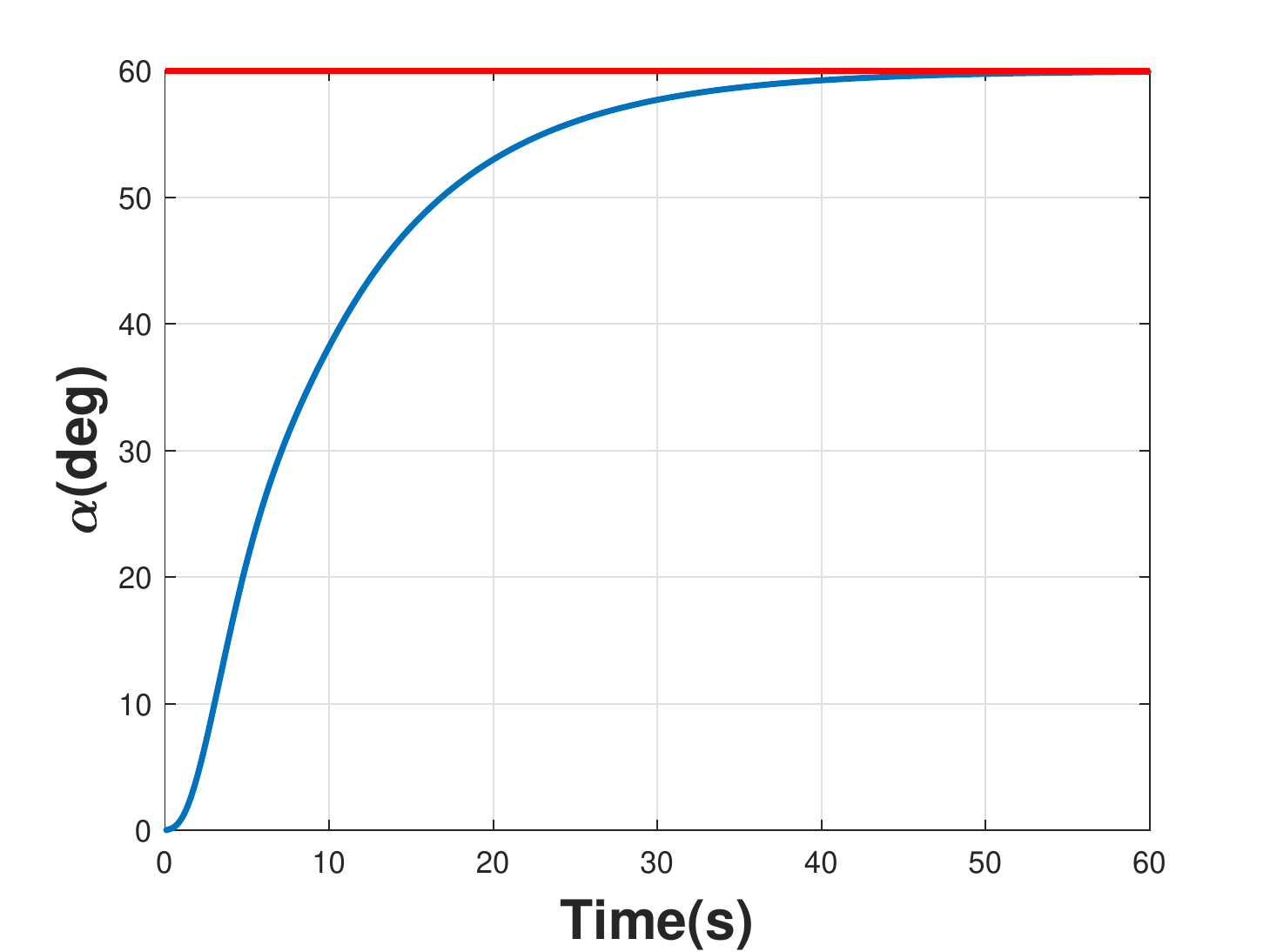}
\caption{\label{fig:alpha} Scenario 1. Tower angle $\alpha$. Red line: Desired reference. Blue line: Simulation result.}
\end{figure}

\begin{figure}[ht!]
\centering
\includegraphics[width=8cm, height=2.5cm]{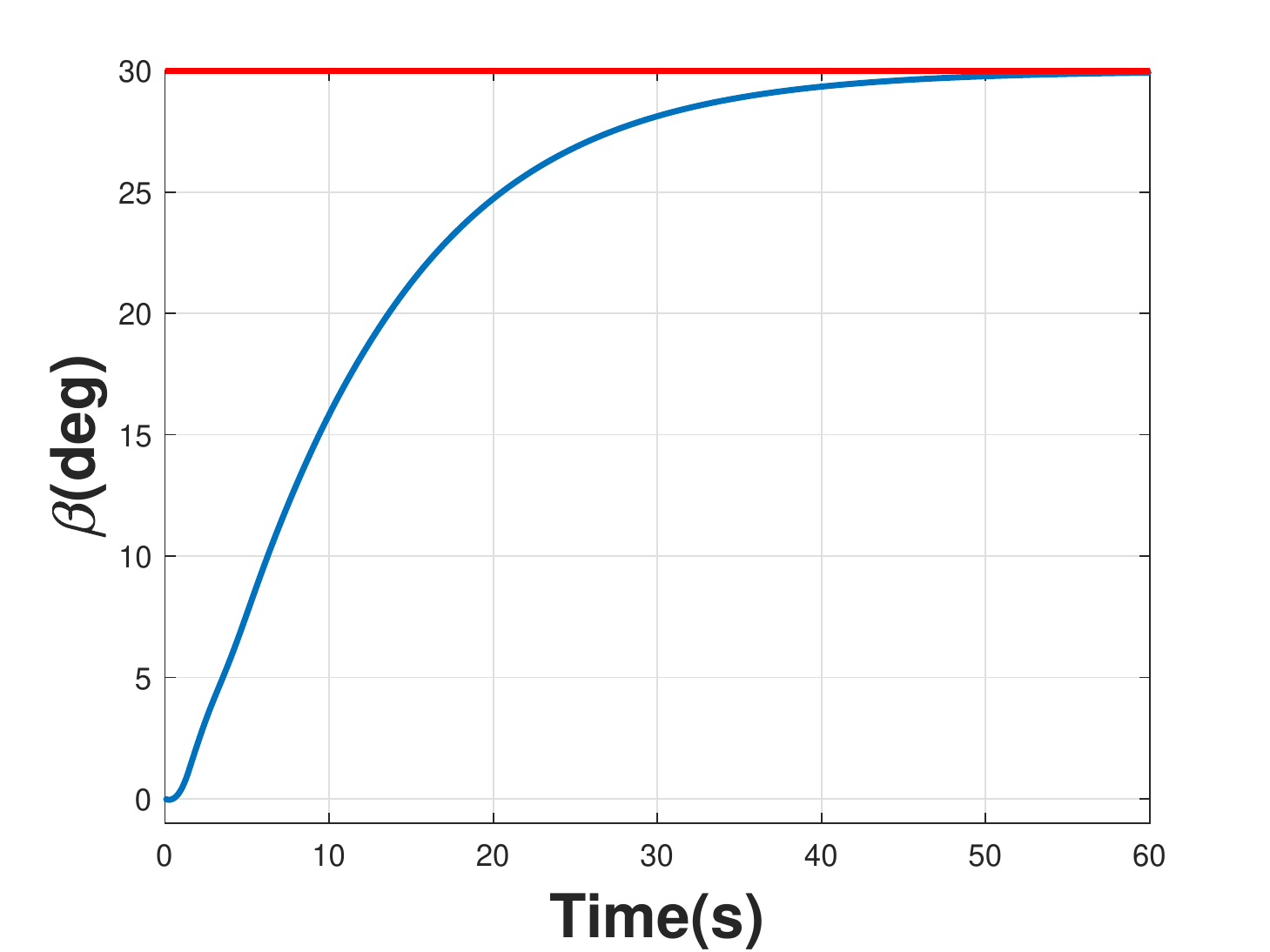}
\caption{\label{fig:beta} Scenario 1. Boom angle $\beta$. Red line: Desired reference. Blue line: Simulation result.}
\end{figure}

\begin{figure}[ht!]
\centering
\includegraphics[width=8cm, height=2.5cm]{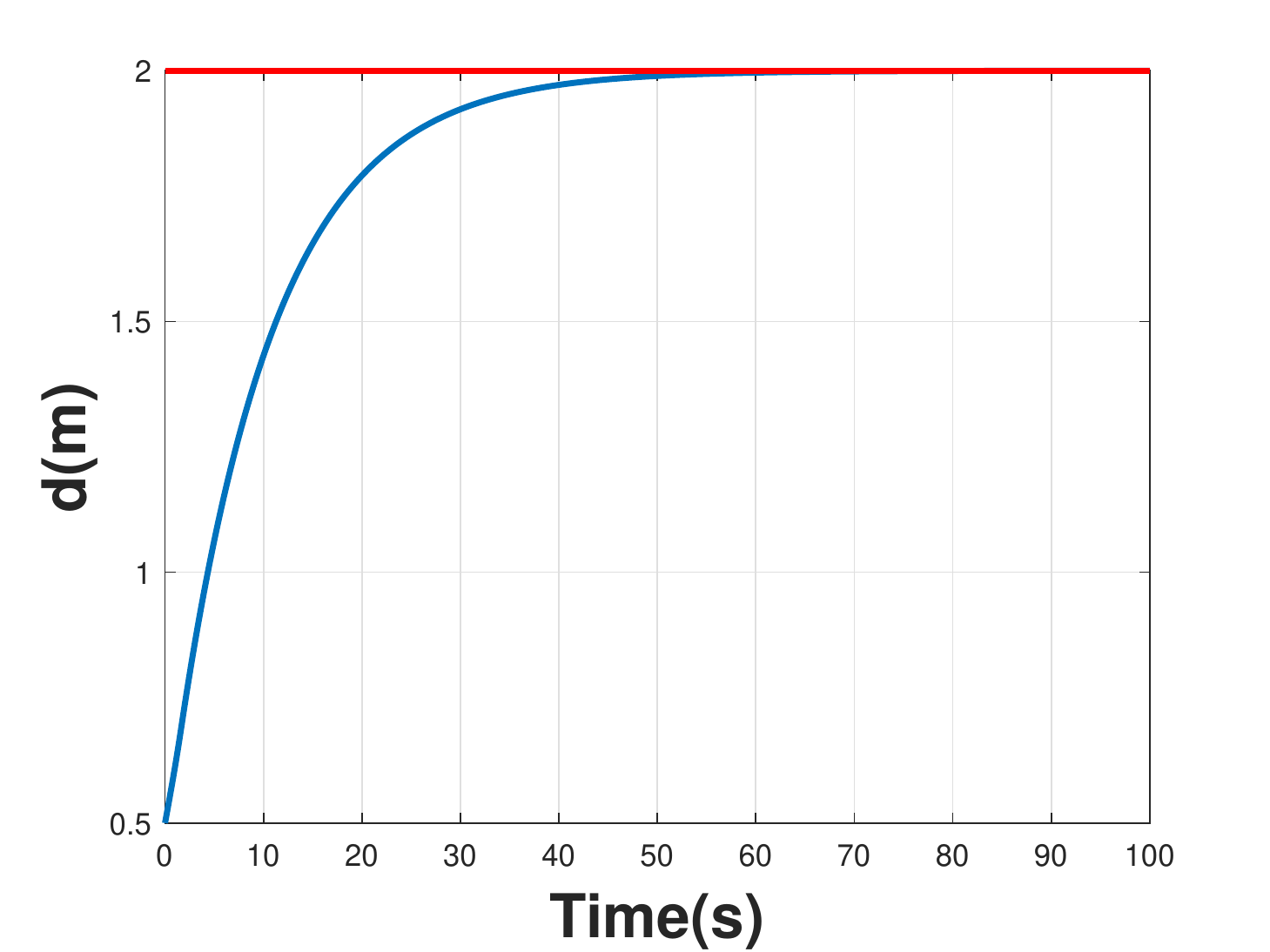}
\caption{\label{fig:d} Scenario 1. Cable length. Red line: Desired reference. Blue line: Simulation result. }
\end{figure}

\begin{figure}[ht!]
\centering
\includegraphics[width=8cm, height=2.5cm]{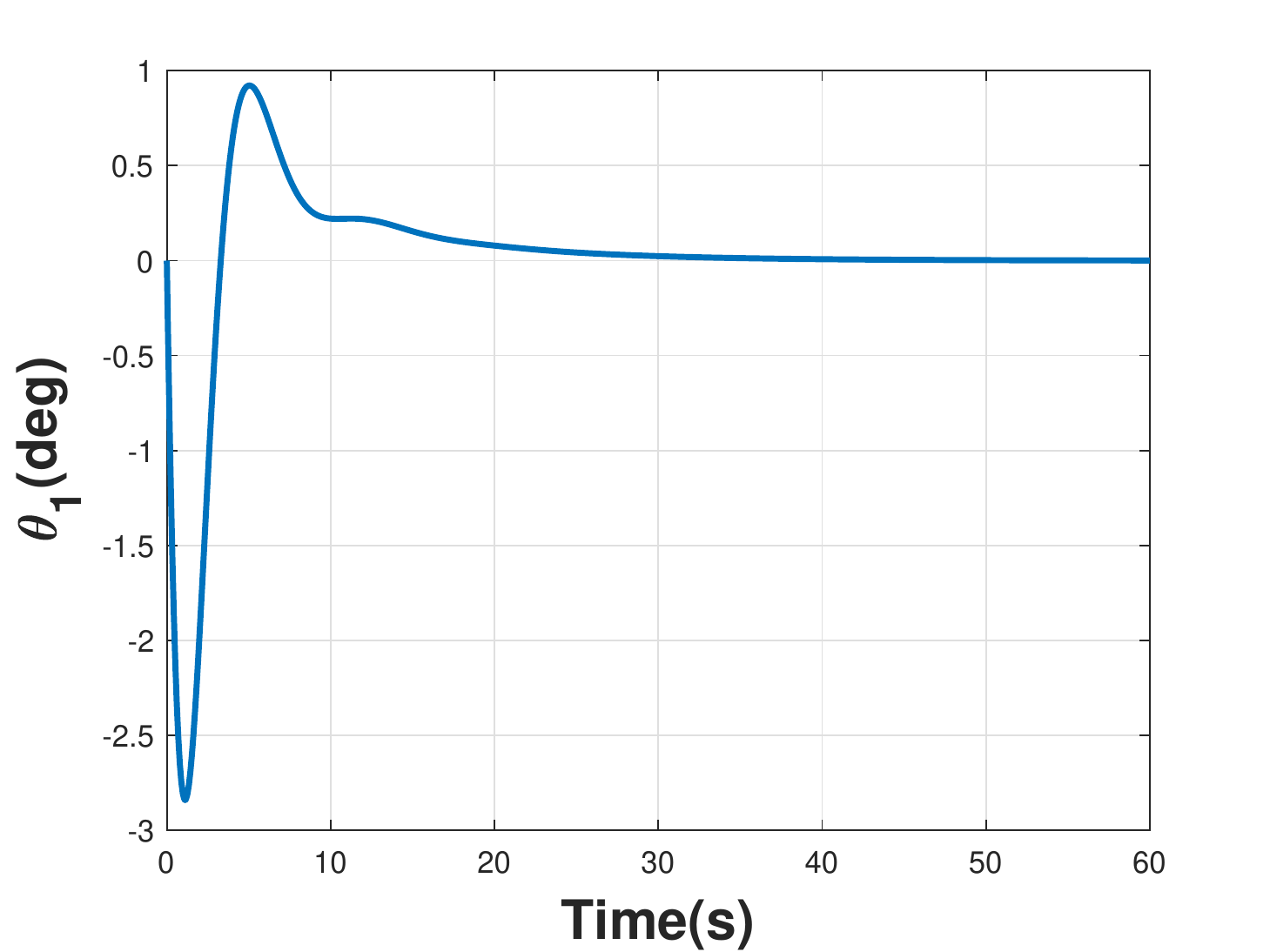}
\caption{\label{fig:th1} Scenario 1. Payload angle  $\theta_1$.}
\end{figure}

\begin{figure}[ht!]
\centering
\includegraphics[width=8cm, height=2.5cm]{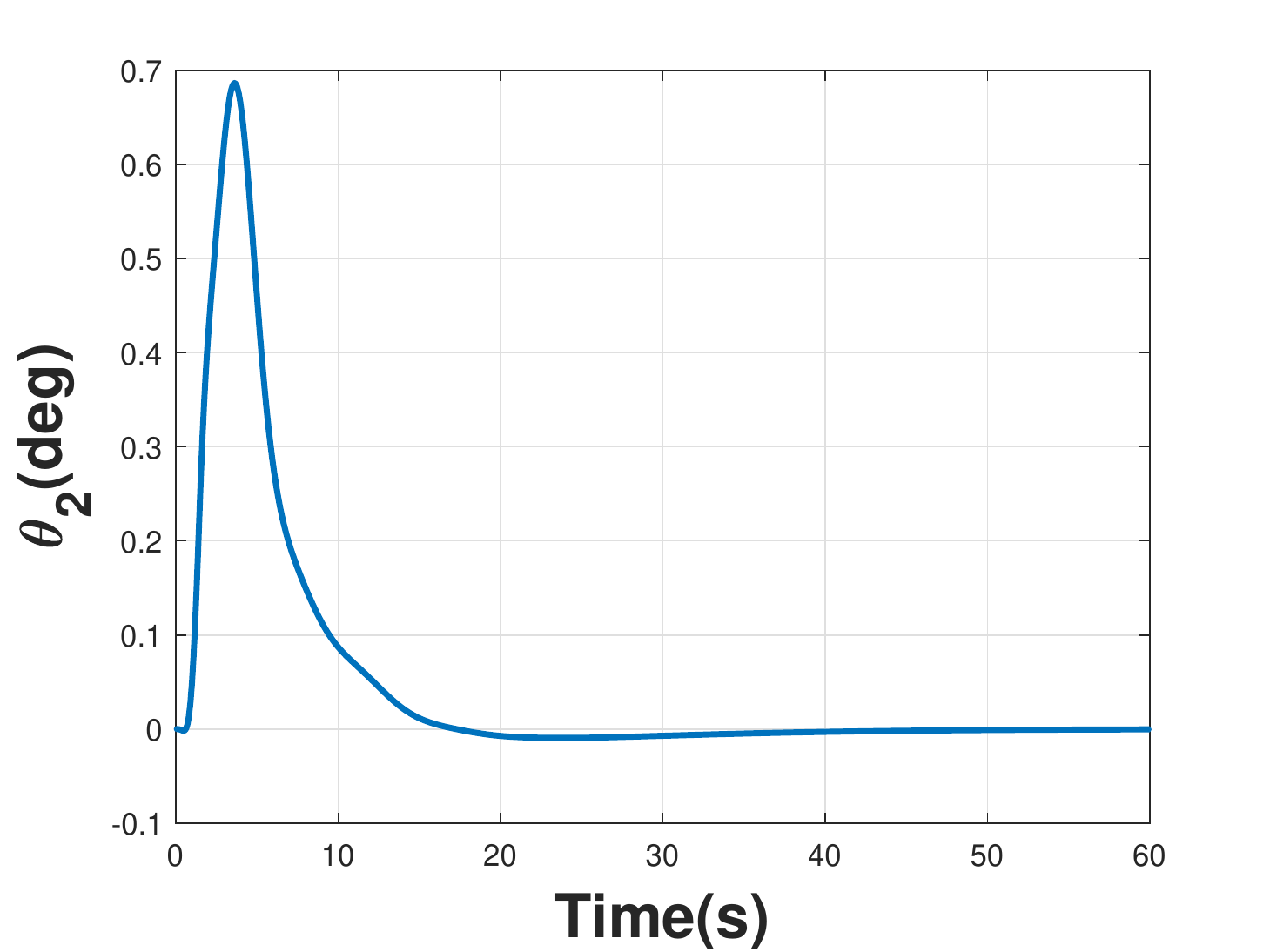}
\caption{\label{fig:th2} Scenario 1. Payload angle  $\theta_2$.}
\end{figure}

\begin{figure}[ht!]
\centering
\includegraphics[width=8cm, height=5cm]{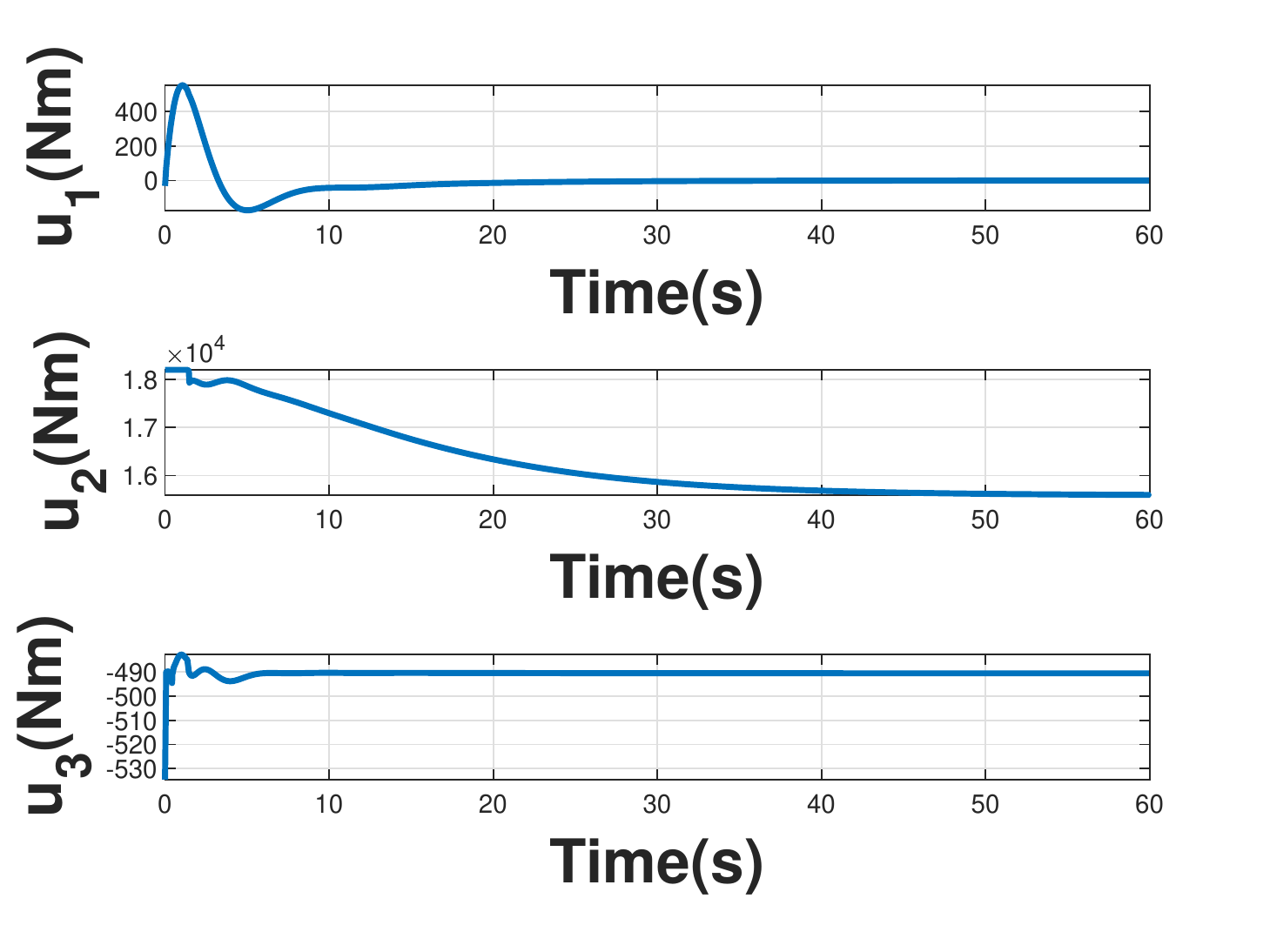}
\caption{\label{fig:u} Scenario 1. Control inputs }
\end{figure}

\textit{Scenario 1.} In this simulation scenario, we show the performance of the proposed control law described in the Section \ref{sec:cntrl}. The goal is to move the crane to a desired configuration while damping the payload swing angles as much as possible. In this first scenario no external disturbances to the crane will be considered. The simulation results are shown in Figg.\ref{fig:alpha}-\ref{fig:th2}. We can see that
the boom arrives at the desired positions in around 30 seconds, Additionally, the maximum payload swing amplitudes in the two directions are confined in $-2.5\degree$ and $1\degree$, respectively. In Fig.\ref{fig:u} the input controls are shown. For the boom actuator following the \cite{NK100_user_manual}, the limit of the working range of the crane is of 210kg for the payload  mass  with  a  boom  length  of  8.9m  then  the  maximal torque should be around $u_{2max}=18.2kNm$. The values of the other two inputs do not represent a problem as the inputs values are reasonable and well within the typical limits of the crane actuators.

\begin{figure}[ht!]
\centering
\includegraphics[width=8cm, height=2.5cm]{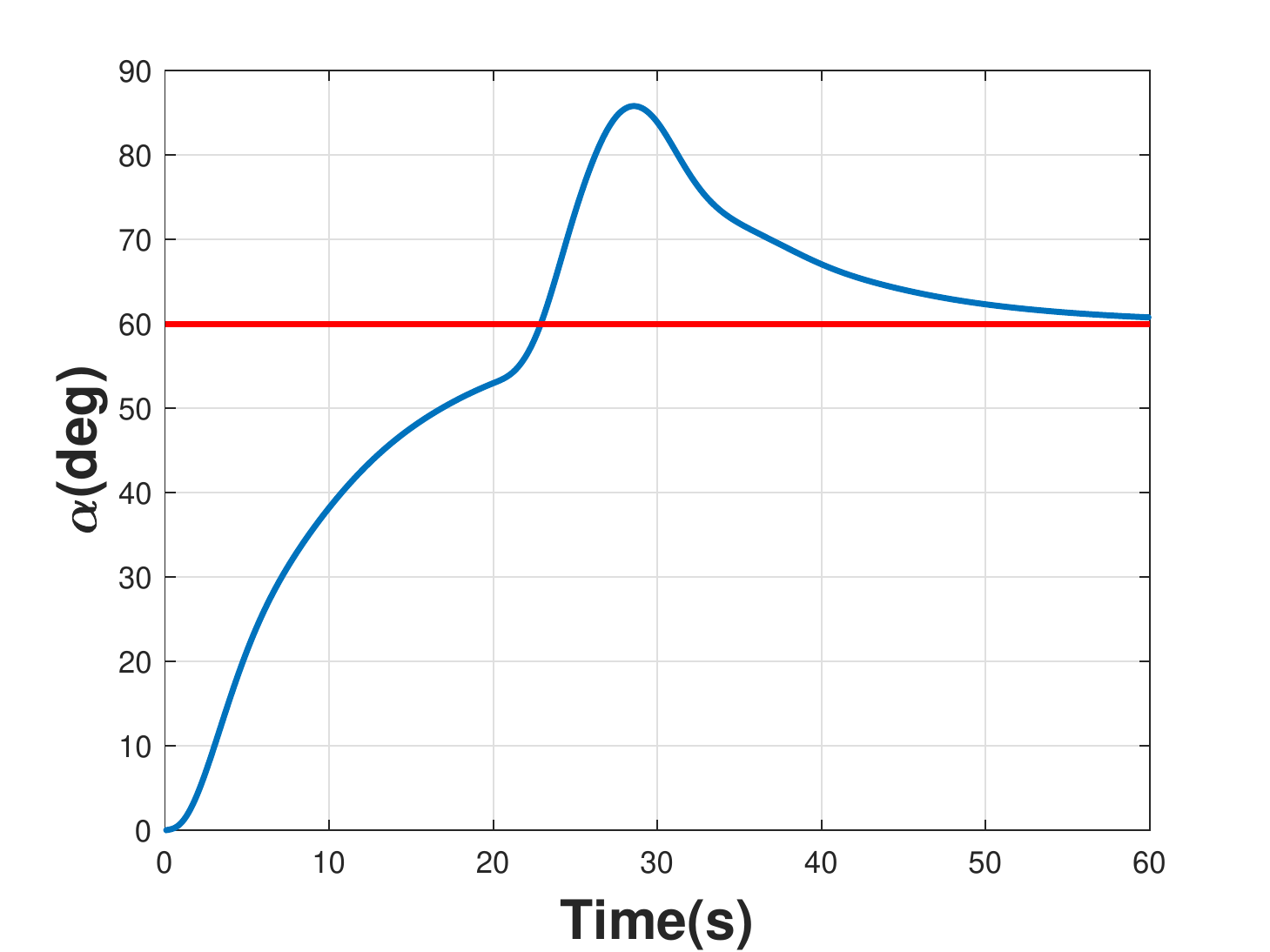}
\caption{\label{fig:alpha_1} Scenario 2. Tower angle $\alpha$. Red line: Desired reference. Blue line: Simulation result.}
\end{figure}

\begin{figure}[ht!]
\centering
\includegraphics[width=8cm, height=2.5cm]{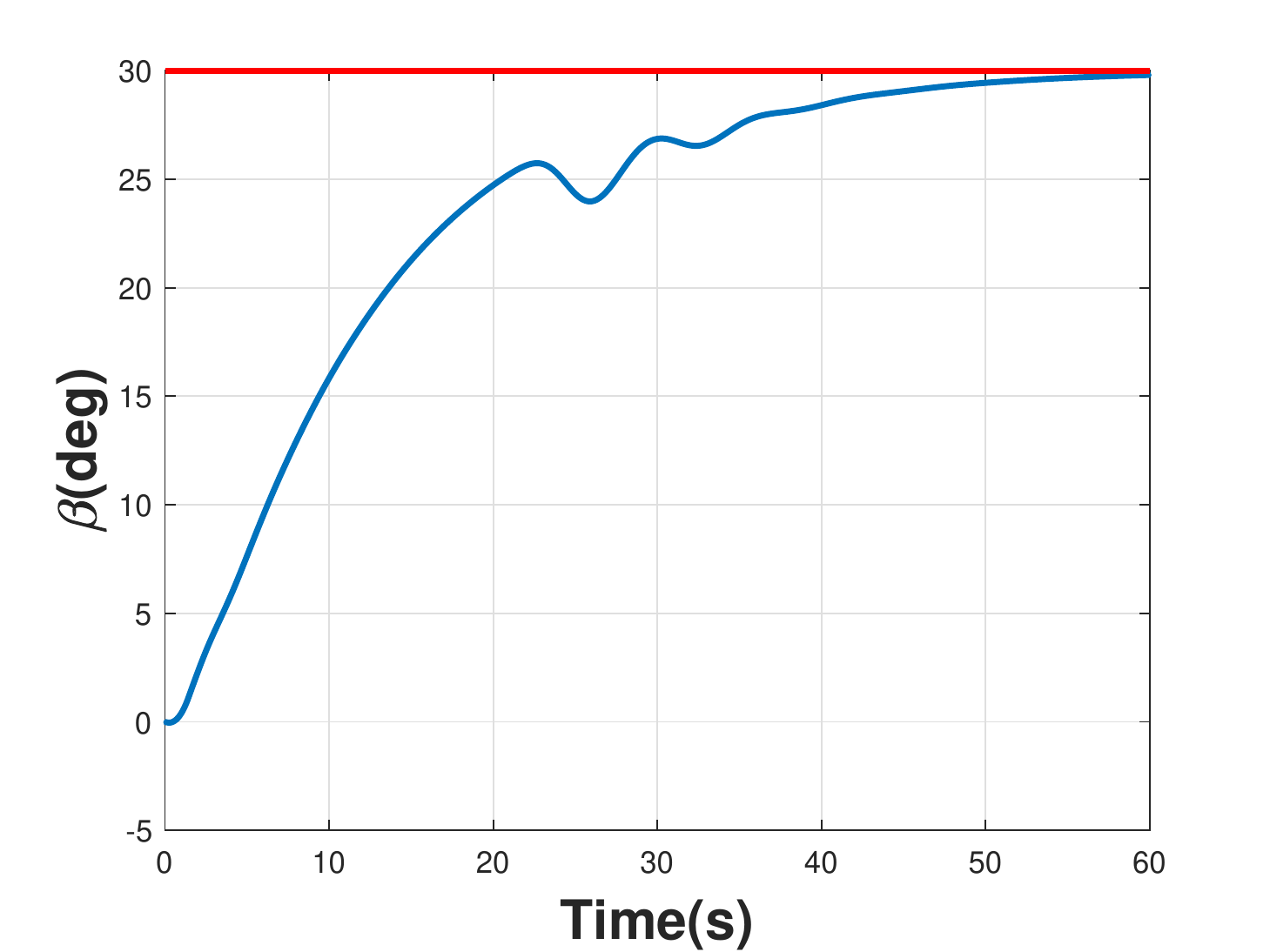}
\caption{\label{fig:beta_1} Scenario 2. Boom angle $\beta$. Red line: Desired reference. Blue line: Simulation result.}
\end{figure}

\begin{figure}[ht!]
\centering
\includegraphics[width=8cm, height=2.5cm]{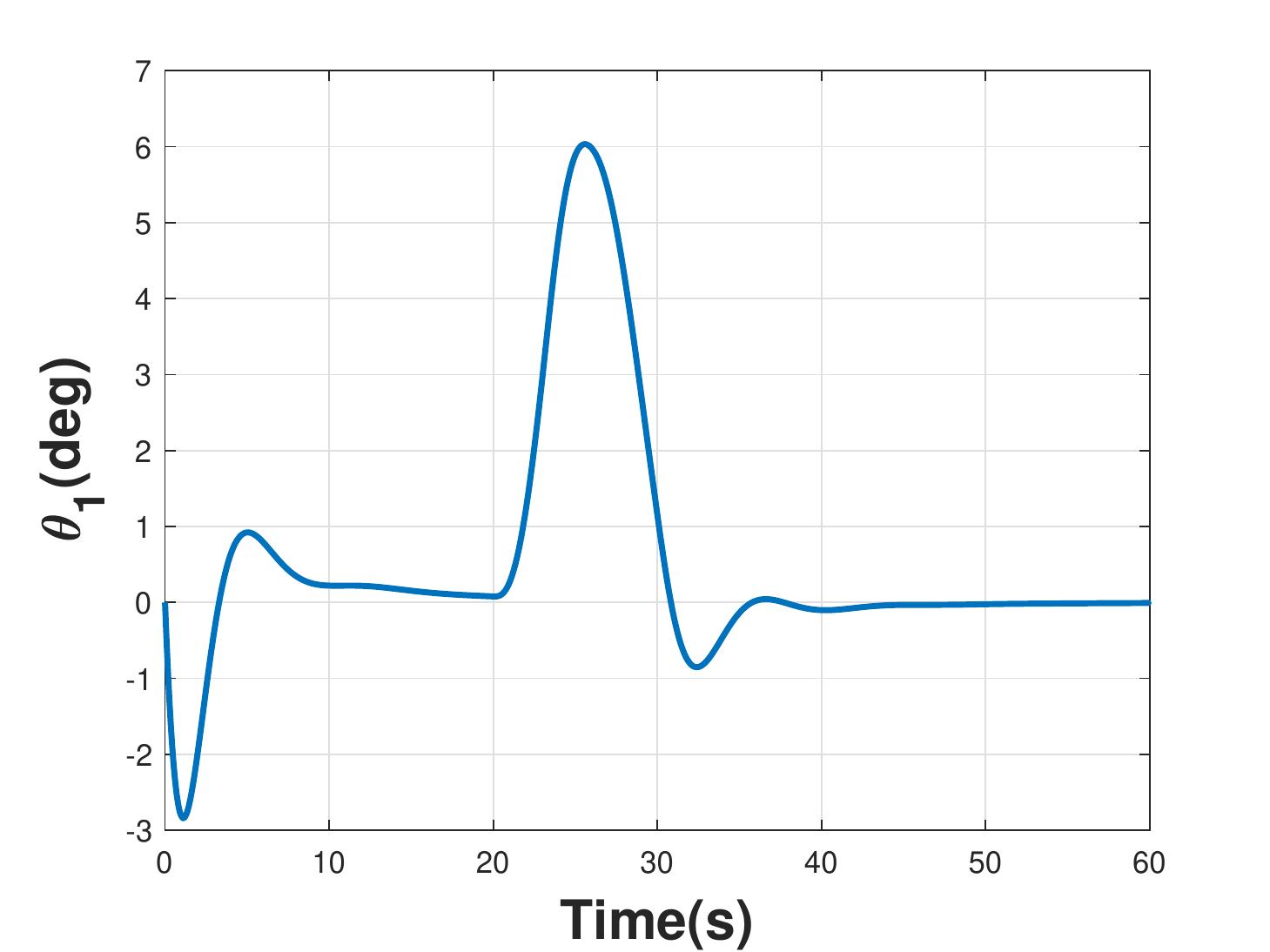}
\caption{\label{fig:th1_1} Scenario 2. Payload angle  $\theta_1$.}
\end{figure}

\begin{figure}[ht!]
\centering
\includegraphics[width=8cm, height=2.5cm]{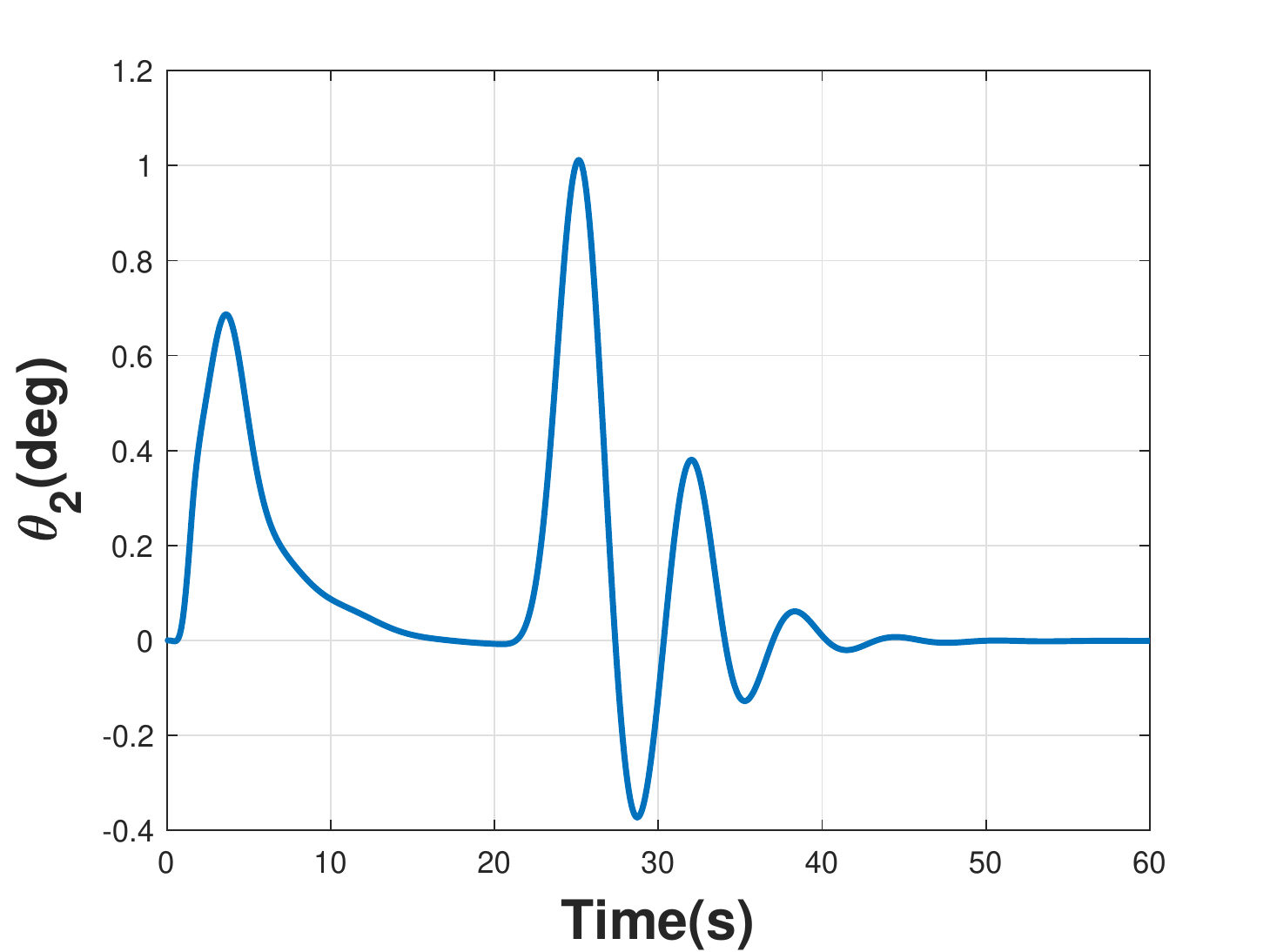}
\caption{\label{fig:th2_1} Scenario 2. Payload angle  $\theta_2$.}
\end{figure}

\begin{figure}[ht!]
\centering
\includegraphics[width=8cm, height=5cm]{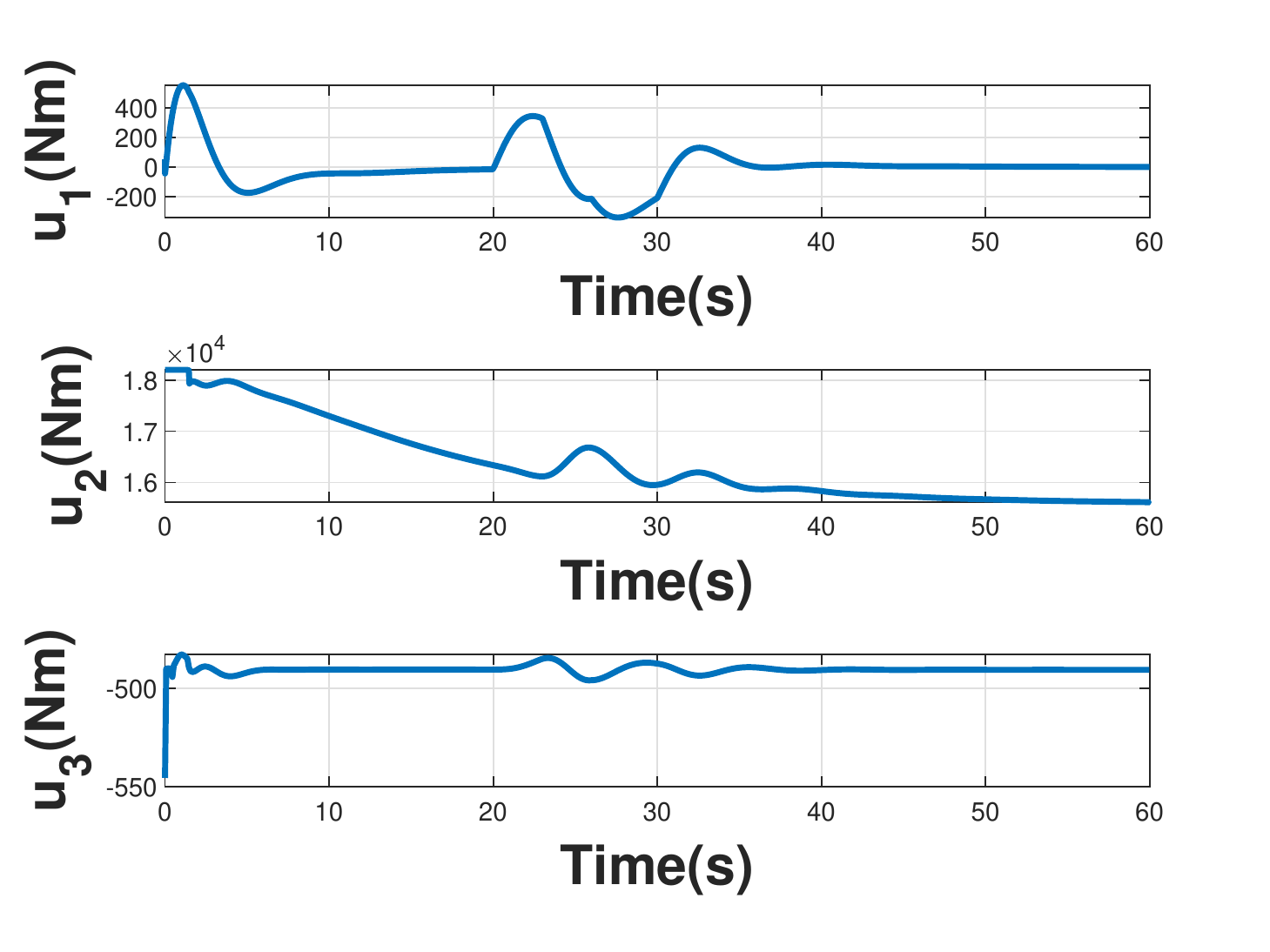}
\caption{\label{fig:u1} Scenario 2. Control inputs }
\end{figure}

\textit{Scenario 2.} In this simulation scenario, we consider a gust of wind as external disturbance for the crane. The desired configuration for the crane is the same of the previous scenario. In this case the controller must be able to counteract the effect of wind during the whole movement of the crane.  The perturbation seen by the system will be characterized by a duration and a time dependent amplitude. Concerning the first one, a study from a meteorological center of the Netherlands reported that wind gusts have periods of 2 to 7 seconds with average speeds comprised between 4 and 20 $m/s$ \cite{verheij_gust_1992}. The force applied on the payload can be seen as distributed force $F = \frac{1}{2}\rho V^2 A_w C_D $, where V is the wind average gust speeds, and $A_w$ is the surface exposed to the wind. According to \cite{wind_influence}, $C_D = 1.05$ will be chosen. Assuming ISA conditions at sea level, $\rho = 1.225 [kg/m^3]$.
\newline
In this Scenario, we consider a force that acts laterally to the load (e.g. increases the swing angle $\theta_1$). In this scenario, only one gust of wind will occur when the simulation is at 20s. In our simulations, the wind speed will have a trapezoidal shape (e.g. increase linearly from zero, constant for a time window and finally linearly decrease to zero). 
\newline
As one can see in Figg.~\ref{fig:th1_1}-\ref{fig:th2_1}, due to the wind gust, the swing angle $\theta_1$ increases and consequently also the angle $\theta_2$ oscillates. To counteract this effect, the controller modifies the tower angle $\alpha$ (Fig.\ref{fig:alpha_1} and Fig.\ref{fig:u1}) and the boom angle $\beta$ (Fig.\ref{fig:beta_1} and Fig.\ref{fig:u1}) to reduce the swing angles as fast as possible. The small effects on the length of the cable can be seen in Fig.\ref{fig:u1}, where one can see that the force on the cable changes a little.    

\begin{figure}[ht!]
\centering
\includegraphics[width=8cm, height=2.5cm]{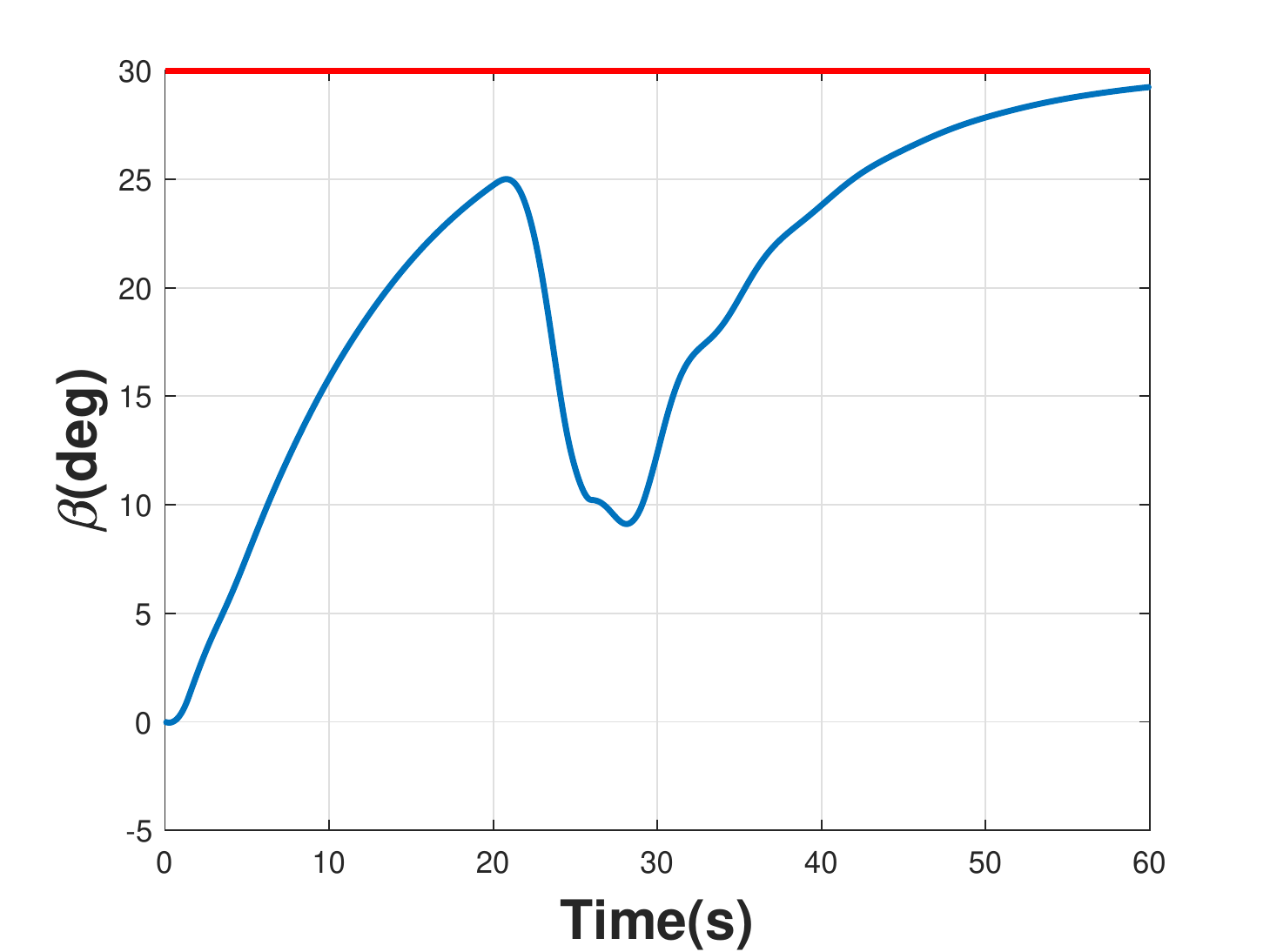}
\caption{\label{fig:beta_2} Scenario 3. Boom angle $\beta$. Red line: Desired reference. Blue line: Simulation result.}
\end{figure}

\begin{figure}[ht!]
\centering
\includegraphics[width=8cm, height=2.5cm]{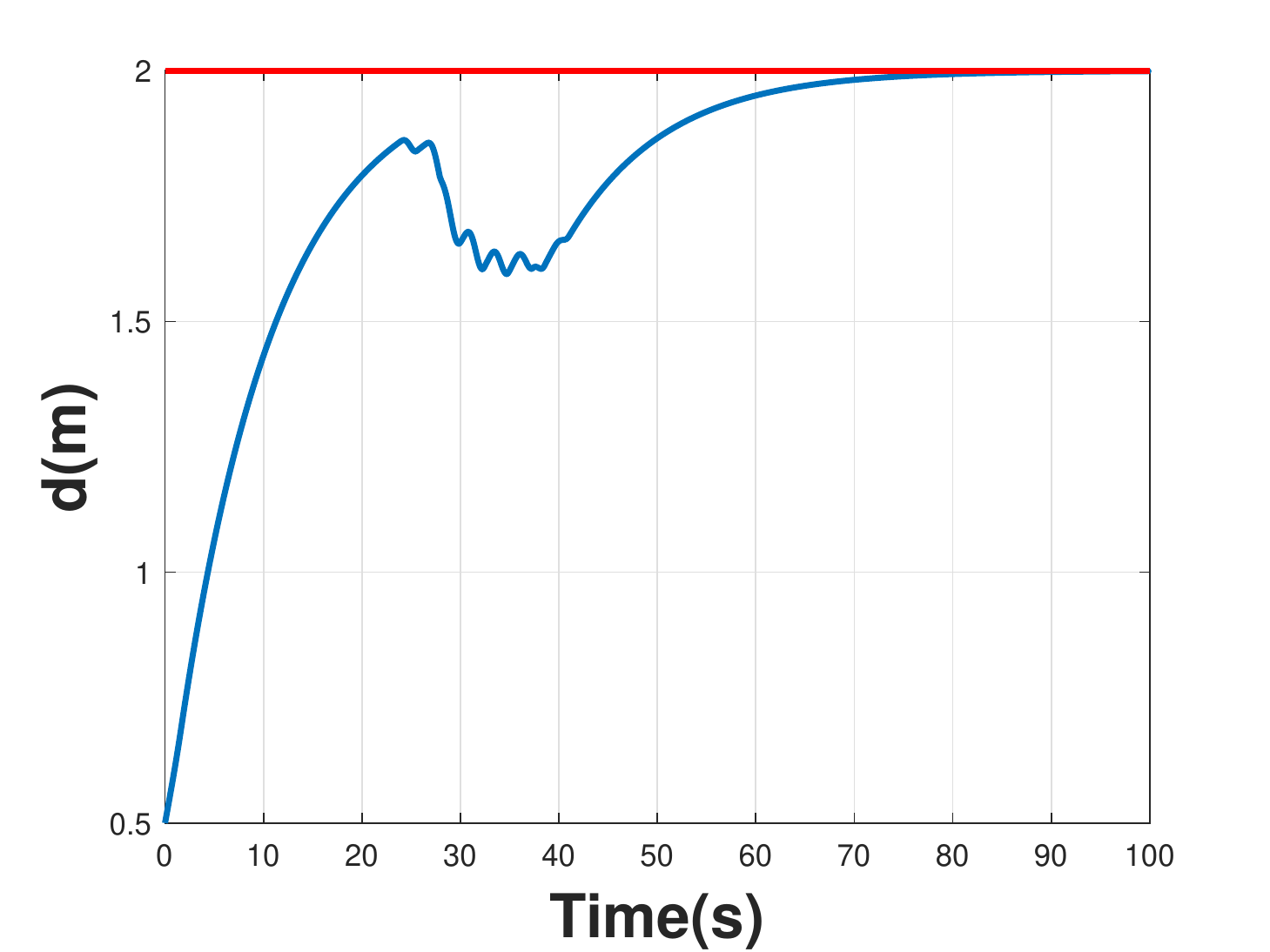}
\caption{\label{fig:d_2} Scenario 3. Cable length. Red line: Desired reference. Blue line: Simulation result. }
\end{figure}

\begin{figure}[ht!]
\centering
\includegraphics[width=8cm, height=2.5cm]{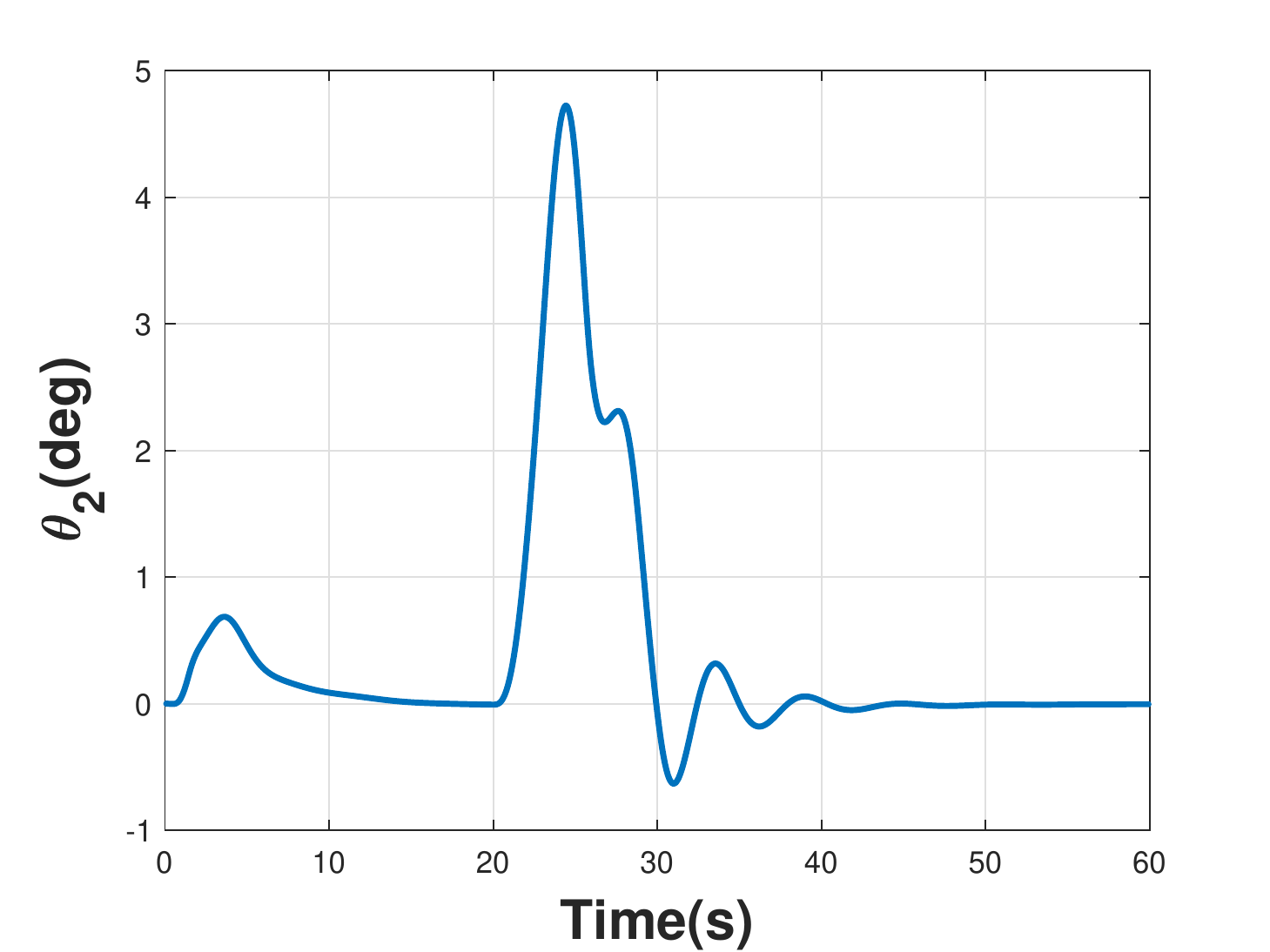}
\caption{\label{fig:th2_2} Scenario 3. Payload angle  $\theta_2$.}
\end{figure}

\begin{figure}[ht!]
\centering
\includegraphics[width=8cm, height=5cm]{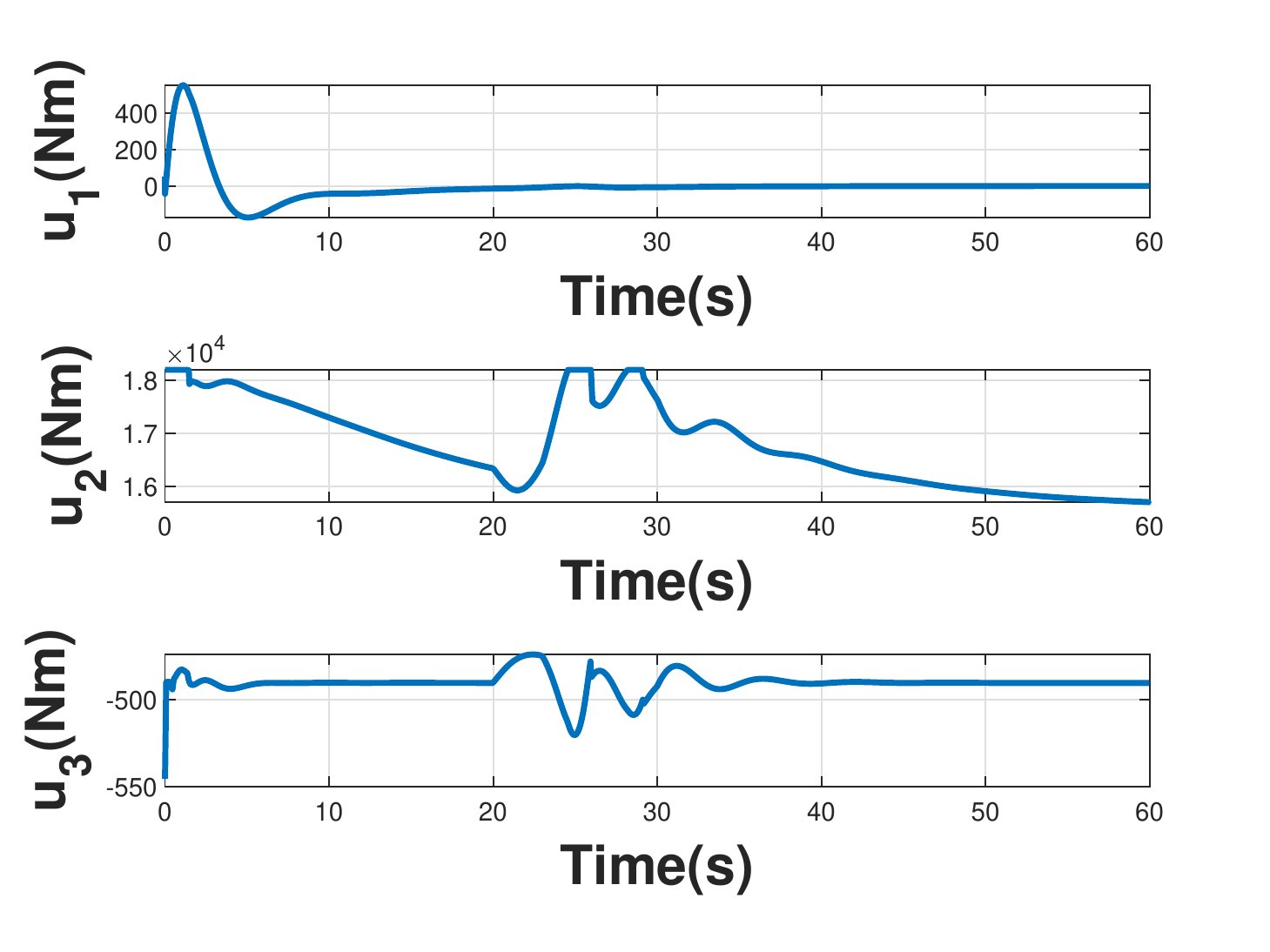}
\caption{\label{fig:u2} Scenario 3. Control inputs }
\end{figure}

\textit{Scenario 3.} In this Simulation scenario, the main effect of the wind is on the angle $\theta_2$. In this case, the swing radial angle increases (see Fig.\ref{fig:th2_2}) and consequentially the controller modifies the value of the luff angle (Fig.\ref{fig:beta_2}) and the length of the cable (Fig.\ref{fig:d_2}) to reduce the oscillations as much as possible. There are no significant effects on the angle $\theta_1$, therefore no changes are required for the slew angle $\alpha$. As one can see in Fig.\ref{fig:u2}, 
to quickly counteract the effect of the wind, the control input $u_2$ reaches its limit value and then decreases.

\begin{figure}[ht!]
\centering
\includegraphics[width=8cm, height=2.5cm]{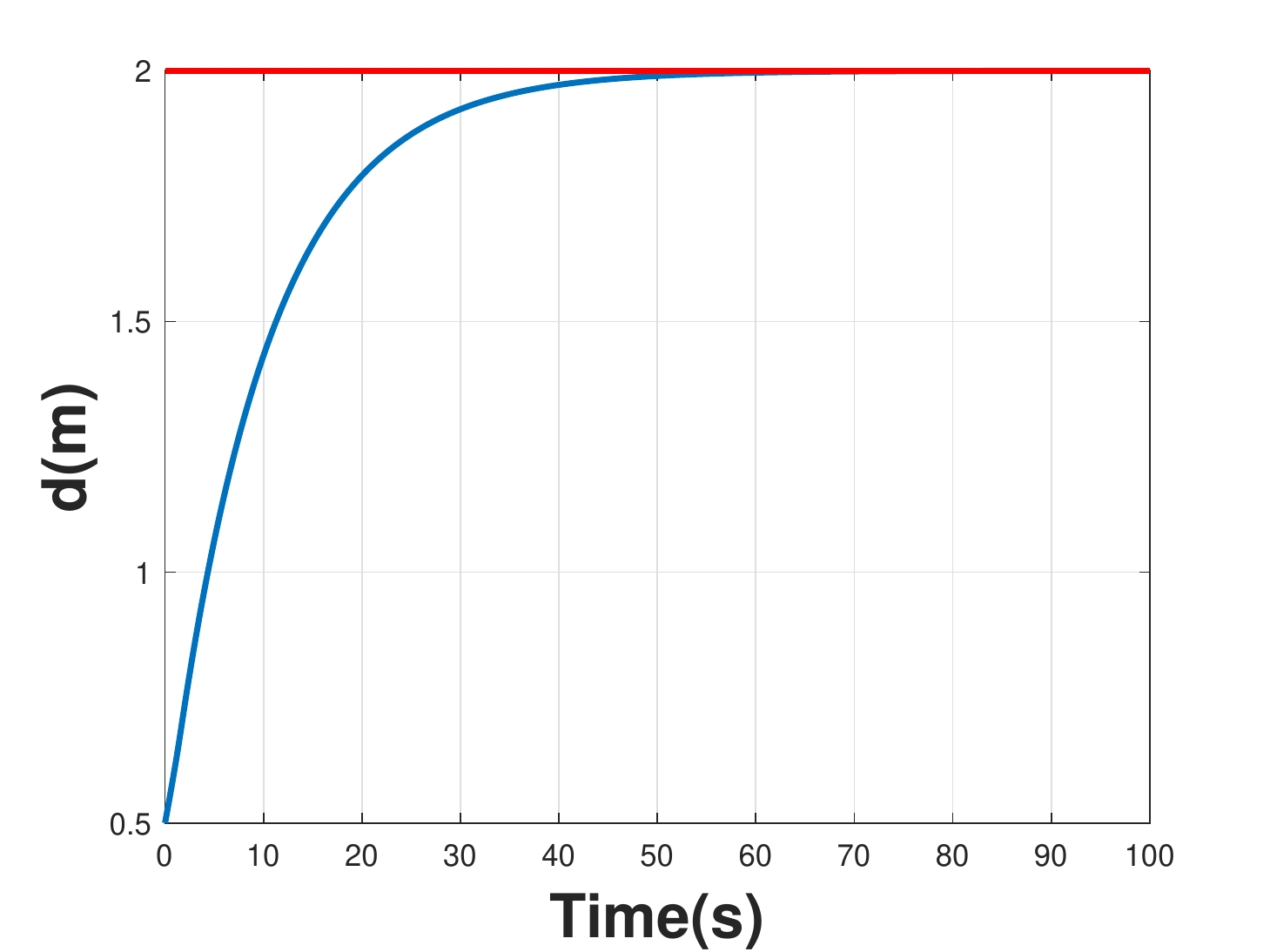}
\caption{\label{fig:d_1} Scenario 2. Cable length. Red line: Desired reference. Blue line: Simulation result. }
\end{figure}

\begin{figure}[ht!]
\centering
\includegraphics[width=8cm, height=2.5cm]{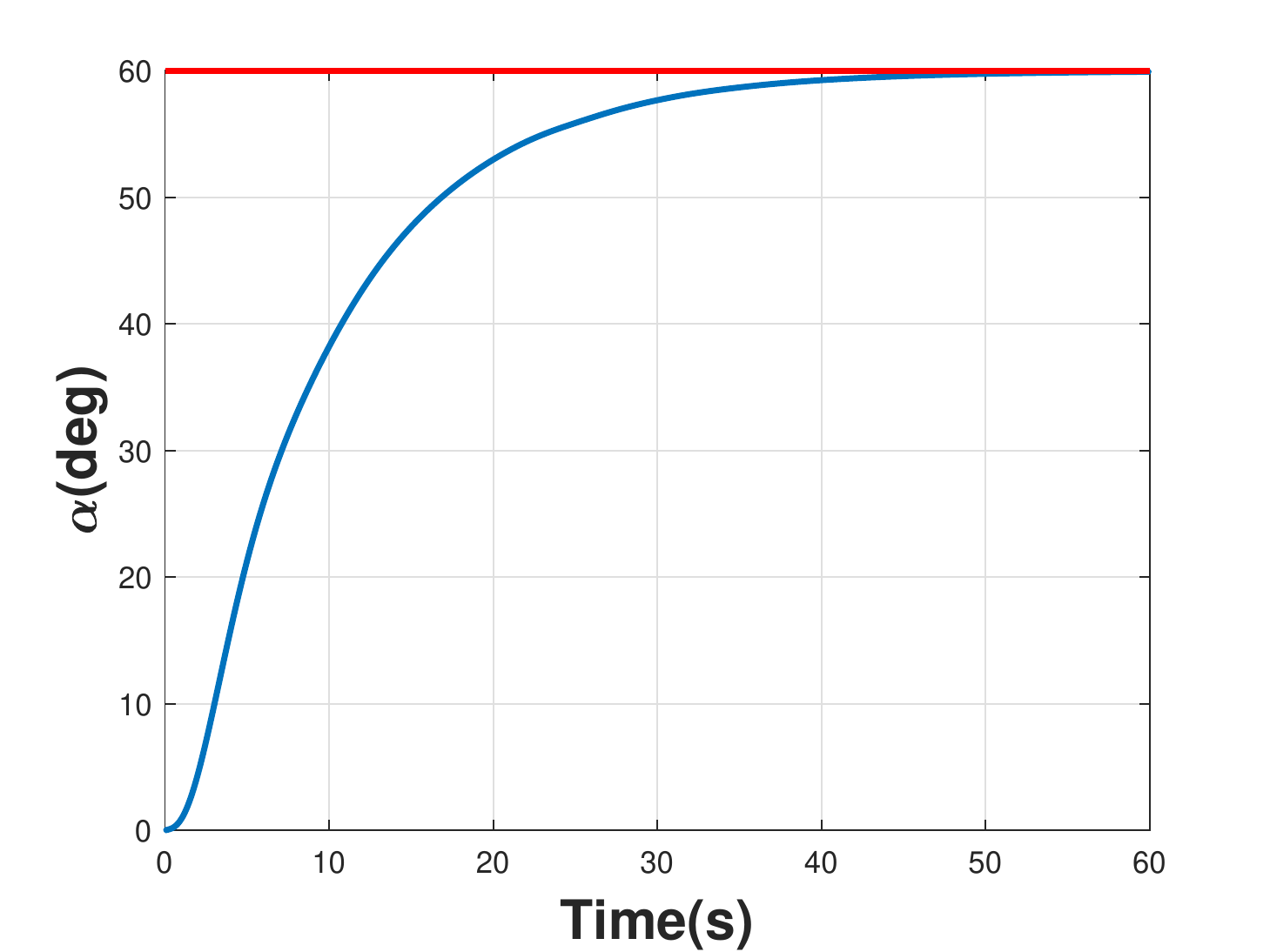}
\caption{\label{fig:alpha_2} Scenario 3. Tower angle $\alpha$. Red line: Desired reference. Blue line: Simulation result.}
\end{figure}

\begin{figure}[ht!]
\centering
\includegraphics[width=8cm, height=2.5cm]{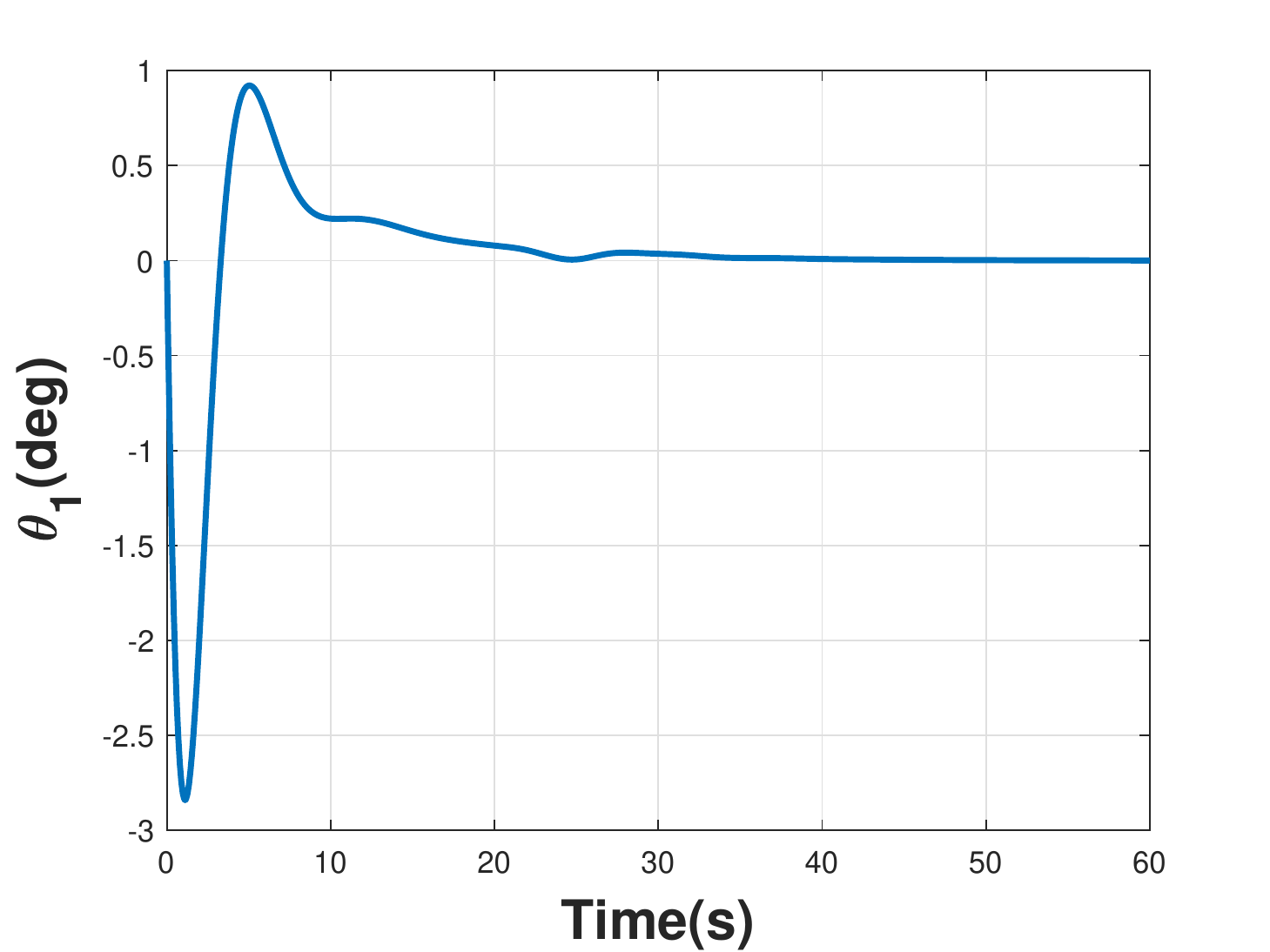}
\caption{\label{fig:th1_2} Scenario 3. Payload angle  $\theta_1$.}
\end{figure}

\section{Conclusion}
\label{sec:Conclusion}
The paper proposed a detailed mathematical model of a boom crane which takes into account all of the degrees of freedom (DoFs) that characterize this type of system (i.e. the two rotations, the length of the rope and the payload swing angles). Despite the complexity of the model, we design a nonlinear control law that exploits all the states of the model to guide the crane towards a desired reference and ensuring that the non-actuated variables (i.e., $\theta_1$ and $\theta_2$) go to zero in a fast way. The simulation results with realistic physical parameters show the efficiency of the proposed control scheme even in the presence of wind disturbance.

\bibliography{ISARC}

\begin{thebibliography}{24}
\providecommand{\natexlab}[1]{#1}
\providecommand{\url}[1]{\texttt{#1}}
\expandafter\ifx\csname urlstyle\endcsname\relax
  \providecommand{\doi}[1]{doi: #1}\else
  \providecommand{\doi}{doi: \begingroup \urlstyle{rm}\Url}\fi

\bibitem[{Tuan} et~al.(2012{\natexlab{a}}){Tuan}, {Kim}, and {Lee}]{25}
L.~A. {Tuan}, G.~{Kim}, and S.~{Lee}.
\newblock Partial feedback linearization control of the three dimensional
  overhead crane.
\newblock pages 1198--1203, 2012{\natexlab{a}}.
\newblock \doi{10.1109/CoASE.2012.6386314}.

\bibitem[{Sun} et~al.(2017){Sun}, {Fang}, {Chen}, and {Lu}]{26}
N.~{Sun}, Y.~{Fang}, H.~{Chen}, and B.~{Lu}.
\newblock Amplitude-saturated nonlinear output feedback antiswing control for
  underactuated cranes with double-pendulum cargo dynamics.
\newblock \emph{IEEE Transactions on Industrial Electronics}, 64\penalty0
  (3):\penalty0 2135--2146, 2017.
\newblock \doi{10.1109/TIE.2016.2623258}.

\bibitem[Tuan et~al.(2014)Tuan, Cuong, Lee, Cong, and Moon]{27}
Le~Tuan, Hoang Cuong, Soon-Geul Lee, Nho Cong, and Kee Moon.
\newblock Nonlinear feedback control of container crane mounted on elastic
  foundation with the flexibility of suspended cable.
\newblock \emph{Journal of Vibration and Control}, 22, 11 2014.
\newblock \doi{10.1177/1077546314558499}.

\bibitem[Sun et~al.(2017{\natexlab{a}})Sun, Fang, Chen, Wu, and lu]{30}
Ning Sun, Yongchun Fang, He~Chen, Yiming Wu, and Biao lu.
\newblock Nonlinear antiswing control of offshore cranes with unknown
  parameters and persistent ship-induced perturbations: Theoretical design and
  hardware experiments.
\newblock \emph{IEEE Transactions on Industrial Electronics}, PP:\penalty0
  1--1, 10 2017{\natexlab{a}}.
\newblock \doi{10.1109/TIE.2017.2767523}.

\bibitem[Uchiyama et~al.(2013)Uchiyama, Ouyang, and Sano]{31}
Naoki Uchiyama, Huimin Ouyang, and Shigenori Sano.
\newblock Simple rotary crane dynamics modeling and open-loop control for
  residual load sway suppression by only horizontal boom motion.
\newblock \emph{Mechatronics}, 23:\penalty0 1223–1236, 12 2013.
\newblock \doi{10.1016/j.mechatronics.2013.09.001}.

\bibitem[Ambrosino et~al.(2020{\natexlab{a}})Ambrosino, Thierens, Dawans, and
  Garone]{kncntr}
M.~Ambrosino, B.~Thierens, A.~Dawans, and E.~Garone.
\newblock Oscillation reduction for knuckle cranes.
\newblock In \emph{ISARC. Proceedings of the International Symposium on
  Automation and Robotics in Construction}, 2020{\natexlab{a}}.

\bibitem[Xi and Hesketh(2010)]{33}
Zhiyu Xi and Tim Hesketh.
\newblock Discrete time integral sliding mode control for overhead crane with
  uncertainties.
\newblock \emph{Control Theory \& Applications, IET}, 4:\penalty0 2071 -- 2081,
  11 2010.
\newblock \doi{10.1049/iet-cta.2009.0558}.

\bibitem[Raja~Ismail and Ha(2013)]{33b}
Raja Mohd~Taufika Raja~Ismail and Quang Ha.
\newblock Trajectory tracking and anti-sway control of three-dimensional
  offshore boom cranes using second-order sliding modes.
\newblock pages 996--1001, 08 2013.
\newblock \doi{10.1109/CoASE.2013.6654071}.

\bibitem[Piazzi and Visioli(2002)]{34}
Aurelio Piazzi and Antonio Visioli.
\newblock Optimal dynamic-inversion-based control of an overhead crane.
\newblock \emph{Control Theory and Applications, IEE Proceedings -},
  149:\penalty0 405 -- 411, 10 2002.
\newblock \doi{10.1049/ip-cta:20020587}.

\bibitem[Sun et~al.(2017{\natexlab{b}})Sun, Wu, Fang, and Chen]{39}
Ning Sun, Yiming Wu, Yongchun Fang, and He~Chen.
\newblock Nonlinear antiswing control for crane systems with double-pendulum
  swing effects and uncertain parameters: Design and experiments.
\newblock \emph{IEEE Transactions on Automation Science and Engineering},
  PP:\penalty0 1--10, 07 2017{\natexlab{b}}.
\newblock \doi{10.1109/TASE.2017.2723539}.

\bibitem[Arnold et~al.(2005)Arnold, Sawodny, Neupert, and Schneider]{44}
Eckhard Arnold, Oliver Sawodny, J.~Neupert, and Klaus Schneider.
\newblock Anti-sway system for boom cranes based on a model predictive control
  approach.
\newblock \emph{IEEE International Conference Mechatronics and Automation,
  2005}, 3:\penalty0 1533--1538 Vol. 3, 2005.

\bibitem[Nakazono et~al.(2008)Nakazono, Ohnishi, Kinjo, and Yamamoto]{45}
Kunihiko Nakazono, Kouhei Ohnishi, Hiroshi Kinjo, and Tetsuhiko Yamamoto.
\newblock Vibration control of load for rotary crane system using neural
  network with ga-based training.
\newblock \emph{Artificial Life and Robotics}, 13\penalty0 (1):\penalty0
  98--101, Dec 2008.

\bibitem[{Uchiyama} et~al.(2012){Uchiyama}, {Ouyang}, and {Sano}]{46}
N.~{Uchiyama}, H.~{Ouyang}, and S.~{Sano}.
\newblock Residual load sway suppression for rotary cranes using only s-curve
  boom horizontal motion.
\newblock pages 6258--6263, 2012.
\newblock \doi{10.1109/ACC.2012.6315369}.

\bibitem[Sano et~al.(2012)Sano, Ouyang, and Uchiyama]{47}
Shigenori Sano, Huimin Ouyang, and Naoki Uchiyama.
\newblock Residual load sway suppression for rotary cranes using simple
  dynamics model and s-curve trajectory.
\newblock \emph{IEEE International Conference on Emerging Technologies and
  Factory Automation, ETFA},
  12818107128151203528138151281510123126851333674122135:\penalty0 1--5, 09
  2012.
\newblock \doi{10.1109/ETFA.2012.6489665}.

\bibitem[Samin et~al.(2014)Samin, Mohamed, Jalani, and Ghazali]{49}
Reza~Ezuan Samin, Zaharuddin Mohamed, Jamaludin Jalani, and Rozaimi Ghazali.
\newblock Input shaping techniques for anti-sway control of a 3-dof rotary
  crane system.
\newblock \emph{Proceedings - 1st International Conference on Artificial
  Intelligence, Modelling and Simulation, AIMS 2013}, pages 184--189, 11 2014.
\newblock \doi{10.1109/AIMS.2013.36}.

\bibitem[Huang et~al.(2013)Huang, Maleki, and Singhose]{51}
Jie Huang, Ehsan Maleki, and W.E. Singhose.
\newblock Dynamics and swing control of mobile boom cranes subject to wind
  disturbances.
\newblock \emph{Control Theory \& Applications, IET}, 7:\penalty0 1187--1195,
  06 2013.
\newblock \doi{10.1049/iet-cta.2012.0957}.

\bibitem[Kondo and Shimahara(2004)]{56}
R.~Kondo and S.~Shimahara.
\newblock Anti-sway control of a rotary crane via switching feedback control.
\newblock 1:\penalty0 748 -- 752 Vol.1, 10 2004.
\newblock \doi{10.1109/CCA.2004.1387303}.

\bibitem[Yang et~al.(2017)Yang, Sun, Qian, and Fang]{yang2017}
Tong Yang, Ning Sun, Yuzhe Qian, and Yongchun Fang.
\newblock An antiswing positioning controller for rotary cranes.
\newblock pages 1586--1590, 07 2017.
\newblock \doi{10.1109/CYBER.2017.8446568}.

\bibitem[Sun et~al.(2017{\natexlab{c}})Sun, Yang, Fang, lu, and Qian]{sun2017}
Ning Sun, Tong Yang, Yongchun Fang, Biao lu, and Yuzhe Qian.
\newblock Nonlinear motion control of underactuated 3-dimensional boom cranes
  with hardware experiments.
\newblock \emph{IEEE Transactions on Industrial Informatics}, PP:\penalty0
  1--1, 09 2017{\natexlab{c}}.
\newblock \doi{10.1109/TII.2017.2754540}.

\bibitem[Ambrosino et~al.(2020{\natexlab{b}})Ambrosino, Dawans, and
  Garone]{erg}
M.~Ambrosino, A.~Dawans, and E.~Garone.
\newblock Constraint control of a boom crane system.
\newblock In \emph{ISARC. Proceedings of the International Symposium on
  Automation and Robotics in Construction}, 2020{\natexlab{b}}.

\bibitem[{Tuan} et~al.(2012{\natexlab{b}}){Tuan}, {Kim}, and {Lee}]{partial}
L.~A. {Tuan}, G.~{Kim}, and S.~{Lee}.
\newblock Partial feedback linearization control of the three dimensional
  overhead crane.
\newblock pages 1198--1203, 2012{\natexlab{b}}.
\newblock ISSN 2161-8089.
\newblock \doi{10.1109/CoASE.2012.6386314}.

\bibitem[NEBOMAT(2005)]{NK100_user_manual}
NEBOMAT.
\newblock \emph{NK 1000 User Manual}.
\newblock NEBOMAT, 2005.

\bibitem[Verheij et~al.(1992)Verheij, Cleijne, and Leene]{verheij_gust_1992}
F.~J. Verheij, J.~W. Cleijne, and J.~A. Leene.
\newblock Gust modelling for wind loading.
\newblock \emph{Journal of Wind Engineering and Industrial Aerodynamics},
  42:\penalty0 947--958, October 1992.
\newblock ISSN 0167-6105.
\newblock URL
  \url{http://www.sciencedirect.com/science/article/pii/016761059290101F}.

\bibitem[Liebherr(2017)]{wind_influence}
Liebherr.
\newblock Wind influence on crane operations, 2017.
\newblock 4th Edition.

\end{thebibliography}

\end{document}